\documentclass[11pt,a4paper]{article}
\usepackage[utf8]{inputenc}
\usepackage{epic,eepic,epsfig,amsmath,amssymb,amsthm,mathabx,longtable}
\usepackage{t1enc,tabularx,subcaption,graphicx,slashbox}
\usepackage{array}
\usepackage{multirow}
\usepackage{blindtext}
\usepackage[francais,english]{babel}
\usepackage{caption}
\usepackage{color}
\usepackage{bbm}
\usepackage{hyperref}

\renewcommand{\geq}{\geqslant}
\renewcommand{\leq}{\leqslant}

\newtheorem{theorem}{Theorem}[section]
\newtheorem{corollary}[theorem]{Corollary}
\newtheorem{proposition}[theorem]{Proposition}

\theoremstyle{remark}
\newtheorem{remark}{Remark}[section]
\theoremstyle{definition}
\newtheorem{definition}{Definition}[section]
\newtheorem*{notation}{Notation}
\theoremstyle{definition}

\theoremstyle{definition}

\newtheorem{assumption}[theorem]{Assumption}

\begin{document}

\title{Spectral influence in networks: An application to Input-Output analysis}

\maketitle

\author{\centerline{Nizar Riane$^\dag$}}

\centerline{$^\dag$ Universit\'e Mohammed V de Rabat, Maroc\footnote{Nizar.Riane@gmail.com} }

\vskip 0.5cm

\begin{abstract}
This paper introduces the concepts of spectral influence and spectral cyclicality, both derived from the largest eigenvalue of a graph's adjacency matrix. These two novel centrality measures capture both diffusion and interdependence from a local and global perspective respectively. We propose a new clustering algorithm that identifies communities with high cyclicality and interdependence, allowing for overlaps. To illustrate our method, we apply it to input-output analysis within the context of the Moroccan economy.
\end{abstract}

\vskip 1cm


\vskip 1cm


\vskip 1cm

\noindent \textbf{Acknowledgements}: The author would like to thank the anonymous referee for their judicious remarks about the first version of this paper, which contributed to significantly improving it.

\vskip 0.5cm
 
\tableofcontents

\vskip 0.5cm

\section{Introduction}

\hskip 0.5cm The concept of centrality in network analysis dates back to early graph theory, with many definitions proposed, each providing unique insights into the connections within a graph. For example, in 1987, Phillip Bonacich \cite{Bonacich1987} introduced a centrality measure of a vertex based on the power centrality of its neighbors, as a way of addressing some of the limitations of other centrality measures, such as degree centrality and eigenvector centrality, where a high score means that a vertex is connected to other vertices that themselves have high scores (\cite{Newman2018} ch.7). Bonacich argued that "in bargaining situations, it is advantageous to be connected to those who have few options; power comes from being connected to those who are powerless. Being connected to powerful others who have many potential trading partners reduces one's bargaining power" (\cite{Bonacich1987}).

Linton Freeman objected to the multiplicity of centrality measures and opposed the introduction of new measures, he considered that the development ofmeasures should clarify the concept, but the opposite effect seems to be achieved (\cite{Freeman1978}).

We believe that concerns about the proliferation of centrality definitions arise from the confusion created by the term \textit{centrality} itself. It no longer designated a unique measure of \textit{importance} but a field of study regrouping different definitions for different purposes.

In this context, we propose a new definition of centrality, called spectral influence, which is based on the spectral radius of the network's adjacency matrix, which we call the \textbf{spectral influence}. This new concept characterizes the importance of vertices from a diffusion and interdependence perspective: a vertex's score will be higher the more important its role in diffusion, through closed walks, across the network.

The spectral influence is a local measure that induces a new global measure we call the \textbf{spectral cyclicality} which characterizes the cyclicality over the graph. Based on this definition, we develop a cyclicality based clustering algorithm to decompose the graph into communities with the highest diffusion properties.

A concept similar to spectral influence was developed in \cite{Restrepo2006} and was named \textbf{dynamical importance} by the authors. They used perturbation theory to study the relative variation of the spectral radius resulting from the suppression of an edge or a vertex.  However, no importance was given to the notion of cycles, and no link with diffusion through cycles was established.

A cycle is the simplest structure that brings redundant paths in network connectivity and feedback effects in network dynamics (\cite{Fan2021}). Cycles in a network are synonymous with interdependence, redundancy, diffusion, cascade effect, contagion, amplification, and attenuation. Cyclical patterns and phenomena are the subject of investigation in various fields. For example, in biology, cyclicality is studied in the context of epidemic spread \cite{Petermann2004}, while in biochemistry, it is explored as a possible explanation for biochemical oscillations \cite{Goldstein2009}. In ecology, researchers investigate the robustness of food webs to biodiversity loss in the presence of cyclical patterns \cite{Dunne2002}, while in finance, cyclicality is examined for its role in the propagation of shocks within financial systems \cite{Haldane2011}, and so on.

In the field of economics, numerous studies have focused on the importance of network structures in understanding interactions between economic agents. These network-based studies emphasize significance of network structure in determining economic agents interaction outcomes. For example, Acemoglu et al.~\cite{Acemoglu2012} showed that microeconomic idiosyncratic shocks may lead to aggregate fluctuations in the presence of intersectoral asymmetric interconnections, in opposite to Robert Lucas (\cite{Lucas1977}) argument that microeconomic shocks would average out, and thus, would only have negligible aggregate effects, but the authors conclude that in economies with balanced intersectoral network structures, such as cycle, the aggregate volatility decays at a constant rate. In an opposite direction, the authors of \cite{Goldstein2009} expose a different conclusion in the case of banking system, where it is shown that a cycle structure increase risk of domino effect.

This paper offers a new perspective, using the centrality concept to investigate inter-industrial flows. In the Leontief input-output tradition \cite{Leontief1936}, industries produce commodities using commodities, and all outputs are also used as inputs. The input-output table traces industrial flows and economic circuits.

There is a close connection between input-output flows and graph theory: in an economy where two industrial sectors, A and B, exchange commodities and services, a flow from sector A to sector B expresses a dependency of the second one to the former of the amount the flow as pointed by Roland Lantner in~\cite{Lantner1972}. Therefore, it is possible to consider the input-output table as the adjacency matrix of a graph, call it the \textbf{influences graph}, each flow measures the linkage (dependency) between two industrial sectors, and the influence graph represent the complex structure of the industrial web.

More non-obvious conclusions arise when applying the tools of graph theory: how much stronger is the influence of sector A on a third sector C when sector B is exchanging a flow with sector C ? how much is the influence of sector A on the whole industrial tissue ?

Spectral influence provides answers to these questions: the higher the spectral influence of a sector, the more significant its role. One can anticipate the industrial structure to collapse if a sector with high spectral influence stops producing, such key sectors could be seen as a \textbf{propulsive industry}, dragging an entire economy behind.

In this configuration, cycles must receive special attention. An ideal cyclical industrial structure represents a scenario where the production of each sector is amplified or dampened by the subsequent sector, leading to an infinite production momentum. In such situations, new investment efforts are reduced, and turnover increases.

In this worldview, spectral cyclicality serves as the appropriate metric for emphasizing redundancy and feedback effects conveyed through the cycles and highly cyclic components constituting \textbf{industrial complexes}. These components contribute to the creation of the highest added value.

We apply our spectral cyclicality analysis to the Moroccan input-output tables (IOTs)~\cite{IOTsOCDE}. We bring to light the profound mutations of the Moroccan economy. By calculating the spectral influence of each industry we identify the propulsive industries and we highlight the unique cyclical character of the Moroccan interindustrial flows. We conclude by identifying the key cyclicality components of the industries that form the core of the Moroccan economy.

Our paper is organized as follows:

\begin{enumerate}

\item[\emph{i}.] In Section~\ref{Basics of graph theory}, we recall the rudiments of graph topology and spectral theory of non-negative matrix.
		
\item[\emph{ii}.] In Section~\ref{Spectral influence and cyclicality}, we introduce the spectral influence of a vertex using spectral radius, we explore its properties and we introduce the spectral cyclicality as a measure of global diffusion on a graph.
	
\item[\emph{iii}.] In Section~\ref{Spectral influence and cyclicality clustering}, we introduce our diffusion-based clustering algorithm following a divisive/agglomerative approach allowing overlapping.
	
\item[\emph{iv}.] In Section~\ref{Application to input-output analysis}, we apply our technique to the Moroccan input-output tables, from $1995$ to $2018$.
	
\end{enumerate}

\section{Basics of graph theory\label{Basics of graph theory}}

\hskip 0.5cm Throughout this paper, our results concern \textbf{weighted directed graphs} with \textbf{non-negative weights}, those results could be adapted to undirected graphs as well. We refer the reader, for example, to \cite{Gross2018} for a comprehensive presentation of the following definitions.

A \textbf{graph} $G$ is an ordered pair $G = (V, A)$ where $V=\{s_1,\hdots,s_n\}$ is the set of \textbf{vertices} and $A\subseteq V^2$ is the set of unordered pairs of vertices, called \textbf{edges}. When the pairs of the set $A$ are directed, the graph is a \textbf{directed graph} and the pairs are the \textbf{arcs}. The graph~\mbox{$G = (V, A, W)$}, endowed with a weight matrix $W= \left ( w_{ij}\right)_{1\leq i,\, j \leq n}$ is a \textbf{weighted graph}.

A \textbf{walk} in~$G$ is an alternating sequence of vertices and arcs,~\mbox{$s_0, a_1,s_1,\hdots,a_n,s_n$}, where, for~\mbox{$1 \leq i \leq n$}, $a_i=s_{i-1}\, s_i$. The length of such a walk is $n$, which is also equal to the number of arcs. A \textbf{closed walk} is one with the same first and last vertex. A \textbf{path} is a walk in which all vertices are distinct.

A \textbf{trail} is a walk with no repeated arcs. If a closed trail is nontrivial then it is a \textbf{circuit}. A \textbf{cycle} is a nontrivial closed path. A \textbf{Hamiltonian cycle} is a cycle that includes every vertex of $V$ and a \textbf{loop} is an arc from a vertex to itself.

If there is a path (in each direction) between each pair of vertices, the graph is \textbf{strongly connected graph}.

Given a weighted digraph $G = (V, A, W)$ and a walk~\mbox{$\mathcal{W}_{s_0s_n}=s_0, a_1,s_1,\hdots,a_n,s_n$} in $G$ of length $n$, one can define its \textbf{value} $v(\mathcal{W}_{s_0s_n})$ as the product
\begin{align}
\label{eqn:1}
v(\mathcal{W}_{s_0s_n})=w_{01}\cdots w_{n-1\,n}
\end{align}
and its \textbf{geometric mean value} as
\begin{align}
\label{eqn:2}
\sqrt[n]{v(\mathcal{W}_{s_0s_n})}=\sqrt[n]{w_{01}\cdots w_{n-1\,n}}
\end{align}

Hamiltonian cycles play an important role in what follows. Next, we will express a closed walk in terms of Hamiltonian cycles:

\begin{proposition}{\textbf{Closed walk decomposition\label{Closed walk decomposition}}\\}
Every closed walk can be decomposed into a union of Hamiltonian cycles.
\end{proposition}

\begin{proof}{\ }\\
Suppose the closed walk is of length $m$ and contains at least one redundant vertex, otherwise it is Hamiltonian. The sequence between two redundant vertices is a closed walk of some length $k<m$ on which the same argument holds recursively, dropping all those closed walks we end with a Hamiltonian cycle.
\end{proof}

From the set of Hamiltonian cycles of a weighted digraph, we distinguish the dominant cycle:

\begin{definition}[\textbf{Dominant cycle}] $\, $\\
Let $G = (V, A, W)$ be a weighted digraph and let denotes by $\mathcal{C}_{j}^{n_j}$ a Hamiltonian cycle of length $n_j$ starting from $s_j$. We call \textbf{dominant cycle} the Hamiltonian cycle $\mathcal{C}_{\star}^{n_{\star}}$ with the maximum geometric mean value.
\end{definition}

Next, we recall some fundamental results from graph topology and non-negative matrix theory. The reader could consult, for example, \cite{Horn2013} for more details.

\begin{theorem}[\textbf{Perron-Frobenius \cite{Horn2013} ch.8}\label{Perron-Frobenius}]$\, $\\ 
Let us consider a non-negative matrix~$W$, with spectral radius~\mbox{$\rho(W)$}. Then:  
\begin{enumerate}
\item[i.]  $\rho(W)$ is an eigenvalue of~$W$, and there
exists a non-negative, non-zero vector~$\mathbf{v}$ such that: 
$$W\mathbf{v}=\rho(W)\mathbf{v}$$

\item[ii.]~\mbox{$ \displaystyle 
\min_{1\leq i\leq n} \sum_{j=1}^n w_{ij} \leq \rho(W) \leq \max_{1\leq i\leq n} \sum_{j=1}^n w_{ij}$} and \mbox{$ \displaystyle 
\min_{1\leq j\leq n} \sum_{i=1}^n w_{ij} \leq \rho(W) \leq \max_{1\leq j\leq n} \sum_{i=1}^n w_{ij}$}.

\end{enumerate}

Moreover, if $W$ is irreducible (which corresponds to the situation of a strongly connected graph), then 

\begin{enumerate}
\item[i.]~the eigenvalue $\rho(W)>0$ is an algebraically simple\footnote{An eigenvalue is algebraically simple if its algebraic multiplicity is $1$, i.e., it appears exactly once in the characteristic polynomial of a matrix}. 

\item[ii.]~there is a unique positive right eigenvector $\mathbf{v}$ such that $W\mathbf{v}=\rho(W)\mathbf{v}$ and $\displaystyle\sum_{i=1}^n \mathbf{v}_i=1$.

\item[iii.]~there is a unique positive left eigenvector $\mathbf{w}$ such that $\mathbf{w}^TW=\rho(W)\mathbf{w}^T$ and and $\mathbf{w}\cdot \mathbf{v}=1$.\\

\end{enumerate}
\end{theorem}

\begin{corollary}[\textbf{Perron-Frobenius \cite{Horn2013} ch.8}\label{Perron-Frobenius corollary}]$\, $\\ 
If $\mathbf{A}$ is a non-negative irreducible matrix and $\mathbf{B}$ is any principal square submatrix of $\mathbf{A}$, then \mbox{$\rho(\mathbf{B})<\rho(\mathbf{A})$}.
\end{corollary}

\begin{theorem}[\textbf{\cite{Horn2013} p.539\label{Traceasymptotic}}] $\, $\\
Let $\mathbf{A}$ be a non-negative matrix and $\text{Tr}$ the trace operator. Then $\rho(\mathbf{A})=\displaystyle\underset{m\rightarrow +\infty}{\lim\sup} \left(\text{Tr}\left(\mathbf{A}^{m} \right)\right)^{\frac{1}{m}}$. The limit superior is replaced with the limit if the matrix is irreducible and primitive (positive at some matrix power $k\geq 1$).
\end{theorem}

\section{Spectral influence and cyclicality\label{Spectral influence and cyclicality}}

\hskip 0.5cm In the sequel, $G=(V,A,W)$ denotes a weighted digraph with a non-negative adjacency matrix $W$.

\subsection{Spectral influence: a local approach}

\begin{notation}{\ }
\begin{itemize}
\item Given a subset $S\subset V$, $W(S)$ designates the adjacency matrix $W$ resulting from the suppression of rows and columns corresponding to the elements of $S$.
\item Given a matrix $\mathbf{A}$, write $\mathbf{A}\geq 0$ if $\mathbf{A}_{ij}\geq 0$, and $\mathbf{A}>0$ if $\mathbf{A}_{ij}> 0$, for all $i,j$.\\
\end{itemize}
\end{notation}

Next, we will introduce the cornerstone of this work:

\begin{definition}[\textbf{Spectral influence}\label{spectral influence}] $\, $ \\ 
Given a weighted graph~\mbox{$G=(V,A,W)$}, we define the \textbf{spectral influence} of a subset~$S$ of $V$, as
\begin{align}
\label{eqn:3}
\mathcal{S}(S)&=\frac{\rho(W)-\rho(W(S))}{\rho(W)}
\end{align}

\noindent One has $0\leq \mathcal{S}(S)\leq 1$.
\end{definition}

In the particular case of a strongly connected graph, the matrix $W$ is irreducible, it follows from the Perron–Frobenius corollary \ref{Perron-Frobenius corollary} that $0<\mathcal{S}(S)$ since $\rho(W(S))< \rho(W)$.

We present the following fundamental result:

\begin{theorem}[\textbf{Spectral radius, closed walks and dominant cycle\label{Circuitstheorem}}] $\, $ \\
The spectral radius of the matrix $W$ is asymptotically a nondecreasing function of the closed walk values of the graph, in particular, it is an increasing function of the dominant cycle value.
\end{theorem}

\begin{proof}{\ }\\
It follows directly from theorem \ref{Traceasymptotic} that
$$
\rho(W)=\displaystyle\underset{m\rightarrow +\infty}{\lim\sup} \left(\text{Tr}\left(W^{m} \right)\right)^{\frac{1}{m}}=\displaystyle\underset{m\rightarrow +\infty}{\lim\sup} \left(\sum_{j=1}^n\left(W^{m} \right)_{jj}\right)^{\frac{1}{m}}
$$

\noindent There is nothing to prove if $\rho(W)=0$ (no cycles). If $\rho(W)>0$, using closed walk decomposition \ref{Closed walk decomposition}, let designate by $\mathcal{W}^m_{jj}$ an $m$-closed walk starting from $j$, $\mathcal{C}^{n_j}_j$ a $n_j$ Hamiltonian cycle starting from $j$, $\mathcal{C}^{n_{\star}}_{\star}$ the dominant cycle and $n_{\star}$ its length, one has for some $m$ sufficiently large

\begin{align*}
\left(\sum_{j=1}^n\left(W^{m} \right)_{jj}\right)^{\frac{1}{m}}&=\left(\sum_{j=1}^n \sum_{\mathcal{W}^m_{jj}} v\left(\mathcal{W}^m_{jj}\right)\right)^{\frac{1}{m}}\\
&=\displaystyle\left(\sum_{j=1}^n\sum_{\mathcal{W}^m_{jj}} \underset{\mathcal{C}^{n_i}_i\subset \mathcal{W}^m_{jj}}{\prod} v\left(\mathcal{C}_i^{n_i}\right) \right)^{\frac{1}{m}}\\
&=\displaystyle \sqrt[n_{\star}]{v\left(\mathcal{C}^{n_{\star}}_{\star}\right)} \left(\sum_{j=1}^n\sum_{\mathcal{W}^m_{jj}} \underset{\mathcal{C}^{n_i}_i\subset \mathcal{W}^m_{jj}}{\prod} \underbrace{\left(\frac{v\left(\mathcal{C}_i^{n_i}\right)}{{v\left(\mathcal{C}^{n_{\star}}_{\star}\right)}^{\frac{n_i}{n_{\star}}}}\right)}_{\leq1}   \right)^{\frac{1}{m}}
\end{align*}

\noindent Then

$$\displaystyle\underset{m\rightarrow +\infty}{\lim\sup} \left(\sum_{j=1}^n\left(W^{m} \right)_{jj}\right)^{\frac{1}{m}}=\mathcal{O}\left(\sqrt[n_{\star}]{v\left(\mathcal{C}^{n_{\star}}_{\star}\right)}\right)$$

\noindent Since

\begin{align*}
\kappa(W,\star)=\left(\sum_{j=1}^n\sum_{\mathcal{W}^m_{jj}} \underset{\mathcal{C}^{n_i}_i\in \mathcal{W}^m_{jj}}{\Pi} \left(\frac{v\left(\mathcal{C}_i^{n_i}\right)}{{v\left(\mathcal{C}^{n_{\star}}_{\star}\right)}^{\frac{n_i}{n_{\star}}}}\right) \right)^{\frac{1}{m}}&\leq \left(n\max_j \sum_{\mathcal{W}^m_{jj}} \underset{\mathcal{C}^{n_i}_i\in \mathcal{W}^m_{jj}}{\Pi} \left(\frac{v\left(\mathcal{C}_i^{n_i}\right)}{{v\left(\mathcal{C}^{n_{\star}}_{\star}\right)}^{\frac{n_i}{n_{\star}}}}\right) \right)^{\frac{1}{m}}\\
&\leq \left(n\,n^{m-1} \right)^{\frac{1}{m}}\\
&=n
\end{align*}

\noindent where the upper bound is the number of closed walks in a complete graph. If we took $m=k\cdot n_{\star}$, and designate by $\mathcal{W}^{m,\star}_{jj}$ the $m$-closed walk starting from $j$ and composed exclusively from the dominant cycle $\mathcal{C}^{n_{\star}}_{\star}$, we obtain a lower bound

\begin{align*}
\kappa(W,\star)=\left(\sum_{j=1}^n\sum_{\mathcal{W}^m_{jj}} \underset{\mathcal{C}^{n_i}_i\in \mathcal{W}^m_{jj}}{\Pi} \left(\frac{v\left(\mathcal{C}_i^{n_i}\right)}{{v\left(\mathcal{C}^{n_{\star}}_{\star}\right)}^{\frac{n_i}{n_{\star}}}}\right) \right)^{\frac{1}{m}}&= \left(n_{\star}+\sum_{j=1}^n\sum_{\mathcal{W}^m_{jj}\neq \mathcal{W}^{m,\star}_{jj}} \underset{\mathcal{C}^{n_i}_i\in \mathcal{W}^m_{jj}}{\Pi} \left(\frac{v\left(\mathcal{C}_i^{n_i}\right)}{{v\left(\mathcal{C}^{n_{\star}}_{\star}\right)}^{\frac{n_i}{n_{\star}}}}\right) \right)^{\frac{1}{m}}\\
&\geq \left(n_{\star} \right)^{\frac{1}{m}}\\
&\rightarrow 1
\end{align*}
\end{proof}

\begin{remark}{\label{SIsituations}\ }\\
\begin{enumerate}
\item The first direct consequence is that the spectral influence is maximal for vertices whose deletion asymptotically reduces the closed walk supremum value in $G$ the most, which is equivalent to the deletion of Hamiltonian cycles with the highest geometric mean value as a consequence of theorem \ref{Circuitstheorem}: suppose the graph contains $l$ cycles ordered in decreasing geometric mean values \mbox{$\displaystyle \sqrt[n_1]{v\left(\mathcal{C}^{n_1}_{i_1}\right)}\geq\hdots\geq \sqrt[n_k]{v\left(\mathcal{C}^{n_k}_{i_k}\right)} \geq\hdots \geq \sqrt[n_l]{v\left(\mathcal{C}^{n_l}_{i_l}\right)}$}, and suppose the removal of a vertex $s$ induce the suppression of the first $k-1$ Hamiltonian cycles in the order, the spectral influence is thus

\begin{align}
\label{eqn:4}
\mathcal{S}(s)&=1-\mathcal{O}\left( \frac{\sqrt[n_k]{v\left(\mathcal{C}^{n_k}_{i_k}\right)}}{\sqrt[n_1]{v\left(\mathcal{C}^{n_1}_{i_1}\right)}} \right)
\end{align}

\item In the case of \textbf{unweighted graphs}, the sum $\displaystyle\sum_{j=1}^n\left(W^{m} \right)_{jj}$ is simply the number of $m$-closed walks, and the spectral influence of a vertex $s$ is a function of the number of closed walks passing through $s$.

\item If the graph is \textbf{not strongly connected}, then the matrix $W$ is reducible and it is similar to bloc triangular matrix $\displaystyle\left(\begin{matrix}
W_{11} & W_{12} & \hdots & W_{1k}\\
0 & W_{22} & \hdots & W_{2k}\\
\vdots & \ddots & \ddots & \vdots\\
0 & \hdots & 0 & W_{kk}\\
\end{matrix}
 \right)$ (\cite{Horn2013} page 63). The spectral radius is \mbox{$\rho(W)=\displaystyle\max\{\rho(W_{11}),\hdots,\rho(W_{nn})\}$} and the same reasoning holds per bloc, while the spectral influence is null for non dominant blocs.

\item Like other mathematical means, the geometric mean is sensible to extreme values.

\item The multiple configurations of spectral influence can be enumerated: denote by $\mathcal{C}^{n_{i_1}}_{i_1}$ the dominant cycle and $\mathcal{C}^{n_{i_s}}_{i_s}$ the dominant cycle after the suppression of a vertex $s$, there are four situations:

\begin{enumerate}
\item the vertex $s$ does not belongs to the dominant cycle and

\begin{enumerate}
\item $s$ does not belongs to a Hamiltonian cycle crossing the dominant cycle: \mbox{$\kappa(W(s),i_1)=\kappa(W,i_1)$} and \mbox{$\mathcal{S}(s)=0$}.

\item $s$ belongs to a Hamiltonian cycle crossing the dominant cycle: \mbox{$\kappa(W(s),i_1)<\kappa(W,i_1)$} and \mbox{$\mathcal{S}(s)>0$} (irreducibility corollary \ref{Perron-Frobenius corollary}).
\end{enumerate}

\item the vertex $s$ belongs to the dominant cycle and

\begin{enumerate}
\item the vertex $s$ does not belongs to a Hamiltonian cycle crossing $\mathcal{C}^{n_{i_s}}_{i_s}$: \mbox{$\kappa(W(s),i_2)=\kappa(W,i_2)$} and \mbox{$\mathcal{S}(s)>0$}.

\item the vertex $s$ belongs to a Hamiltonian cycle crossing $\mathcal{C}^{n_{i_s}}_{i_s}$: \mbox{$\kappa(W(s),i_2)<\kappa(W,i_2)$} and \mbox{$\mathcal{S}(s)$} is greater than situation (b).i.
\end{enumerate}
\end{enumerate}

\item The spectral influence is a relative local measure of interdependence and diffusion in networks, it measures the relative cyclicality involving the vertex in question.

\end{enumerate}
\end{remark}

\subsection{Spectral influence: a global approach}

\hskip 0.5cm The concept of spectral influence was introduced as a measure of local influence, and it was demonstrated that this quantity depends on closed walks involving the vertex under consideration, which in turn require the existence of Hamiltonian cycles (theorem \ref{Circuitstheorem}). The total sum of the elementary influences will be as large as the number of vertices involved in Hamiltonian cycles, reaching its maximum when all the vertices are contained in one dominant cycle (\textbf{redundancy effect}). Conversely, the total sum will decrease as the flow is divided into smaller Hamiltonian cycles and loops.

Given a subset $S \subseteq V$ of cardinal $k$, one can write
\begin{align}
\label{eqn:5}
0\leq \sum_{s\in S}\left(\rho(W)- \rho(W(s))\right) \leq k \,\rho(W)
\end{align}

In the case of strongly connected graphs: the left hand bound is never reached by Perron–Frobenius corollary \ref{Perron-Frobenius corollary}. Conversely, the right hand bound is strict except the case where the graph is a Hamiltonian  cycle (the suppression of a vertex disconnect the graph). The redundancy effect occurs when a dominant cycle of length $l\geq 2$ contains $l$ vertices whose suppression reduces the spectral radius and increases the spectral influence. This has the following consequence:

\begin{corollary}[\textbf{The whole is not the sum of the parts}\label{Whole}] $\, $ \\ 
Let $G=(V,A,W)$ be a weighted graph and $S$ be a subset of $V$ of cardinal $k\geq 2$. \textbf{The whole is not the sum of the parts} in the sense that, in general
\begin{align}
\label{eqn:6}
\sum_{s\in S}\left(\rho(W)- \rho(W(s))\right)&\neq \rho(W)-\rho(W(S))
\end{align}
\end{corollary}

\begin{proof}{\ }\\
Figure \ref{Fig1} gives a counter-example.

\begin{figure}[!htb]
\begin{center}
\includegraphics[width=0.5\linewidth]{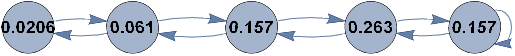}
\end{center}
\caption{The whole is not the sum of the parts: for $S_1=\{1,2\}$ and $S_2=\{4,5\}$, \mbox{$\sum_{s\in S_1}\left(\rho(W)- \rho(W(s))\right)=0.42 < \rho(W)-\rho(W(S_1))=1.74$} while \mbox{$\sum_{s\in S_2}\left(\rho(W)- \rho(W(s))\right)=0.08 > \rho(W)-\rho(W(S_2))=0.06$}.}
\label{Fig1}
\end{figure}
\end{proof}

\begin{remark}{\ \label{UnionRemark}}\\
$\mathcal{S}(S)$ could be interpreted as a measure of a union, i.e., $\mathcal{S}(S)=\mathcal{S}(\bigcup_{s\in S} \{s\})$. Take for example $S=\{i,j\}$:
\begin{align*}
\rho(W)\left(\mathcal{S}(i)+\mathcal{S}(j)\right)&=\rho(W)- \rho(W(i)) + \rho(W)- \rho(W(j))\\
&=\left[\rho(W)- \rho(W(i,j))\right] + \left[\rho(W) - \rho(W(i)) - \rho(W(j)) + \rho(W(i,j))\right] \\
&=\rho(W)\left(\mathcal{S}(i \cup j) + \mathcal{S}(i \cap j) \right)
\end{align*}

\noindent where $\mathcal{S}(i \cap j)=\rho(W) - \rho(W(i)) - \rho(W(j)) + \rho(W(i,j))$. In general, the inclusion-exclusion formula applies.
\end{remark}

According to remark \ref{UnionRemark} above, $\mathcal{S}$ is a measure on $\mathcal{P}(V)$ since
\begin{align}
\label{eqn:6bis}
\mathcal{S}(\emptyset)&=\rho(W)-\rho(W)=0
\end{align}
\noindent and, for any set $S$ of vertices where no two vertices are strongly connected, we have the $\sigma$-additivity
\begin{align}
\label{eqn:6ter}
\mathcal{S}(\bigcup_{s\in S} \{s\})&= \sum_{s\in S} \mathcal{S}(s)
\end{align}

We established the following result:

\begin{proposition}{\ \label{MathematicalMeasure} }\\
Let $G=(V,A,W)$ be a weighted digraph and $\mathcal{P}(V)$ the power set of $V$. $\mathcal{S}$ is a mathematical measure on $\mathcal{P}(V)$.
\end{proposition}

\begin{definition}[\textbf{Spectral cyclicality}\label{spectral cyclicality}] $\, $ \\ 
Given a weighted graph~\mbox{$G=(V,A,W)$}, we define the \textbf{spectral cyclicality} of $G$ as
\begin{align}
\label{eqn:7}
\mathbf{S}(V)&= \displaystyle\sum_{s\in V} \mathcal{S}(s)
\end{align}
\end{definition}

The spectral cyclicality characterizes interdependence and diffusion in the graph. Based on \mbox{equation \ref{eqn:5}}, $0\leq \mathbf{S}\leq n$ and three extreme configurations could occur:

\begin{enumerate}
\item \textbf{The situation of perfect cyclicality : $\mathbf{S}=n$}

\begin{itemize}
\item \textbf{Hamiltonian cycle}: when the connection structure is reduced to a Hamiltonian cycle and the elementary spectral influence is $1$ (the destruction of a vertex disconnect the graph). The spectral cyclicality is $n$.
\end{itemize}

\item \textbf{Situations of weak cyclicality : $\mathbf{S}=1$}

\begin{itemize}
\item \textbf{Fair division}: each vertex uniformly shares the weight $w_i$ with all the vertices. The spectral radius of the matrix equals the mean flow $\displaystyle\frac{\sum_{i=1}^n w_i}{n}$ and the spectral influence is 
$$
\mathcal{S}(s)=\frac{\displaystyle\frac{\sum_{i=1}^n w_i}{n}-\displaystyle\frac{\sum_{i\neq s} w_i}{n}}{\displaystyle\frac{\sum_{i=1}^n w_i}{n}}=\frac{w_s}{\sum_{i=1}^n w_i}
$$
The spectral cyclicality is $1$.

\item \textbf{Loop}: the unique cycle is a loop. The spectral influence equals one for the looped vertex and null elsewhere and the spectral cyclicality is $1$.
\end{itemize}

\item \textbf{No cycles : $\mathbf{S}=0$}

\begin{itemize}
\item \textbf{Chain}: each vertex $s_i$ is connected to the next vertex $s_{i+1}$ constituting a chain.

\end{itemize}

\end{enumerate}

Following remark \ref{SIsituations}, one can deduce that the spectral cyclicality is greater as much as the graph dominant cycle is longer and the graph is strongly connected, and the maximum is reached when the graph is a Hamiltonian cycle. Conversely, spectral cyclicality is reduced by the introduction of non-dominant cycles with high geometric mean values, disjoint from the dominant one, reducing the vertices spectral influence by reducing dominant cycles gap and redundancy effect.

\begin{remark}{\ }\\
From remark \ref{UnionRemark} and proposition \ref{MathematicalMeasure}, the spectral cyclicality reflects redundancy as the sum of the measures increases with the number of common cycles between the vertices.
\end{remark}

Let $G=(V,A,W)$ be a graph and $W(S)$ the weight matrix resulting from the suppression of the vertices subset $S$. In the spirit of Davis-Kahan theorem \cite{Davis-Kahan1970} which quantifies the difference between the subspaces spanned by the eigenvectors of a perturbed symmetric matrix and the original one, we will quantify and interpret the difference \mbox{$\mathbf{S}(V\setminus S)-\mathbf{S}(V)$}, which will provide the foundational motivation of our clustering algorithm. Let consider a perturbation in the sense of a suppression of a subset $S$ of the vertices.
\begin{align*}
\mathbf{S}(V\setminus S)-\mathbf{S}(V)&=\displaystyle \sum_{s\in V\setminus S} \mathcal{S}_{V\setminus S}(s) - \sum_{s\in V} \mathcal{S}(s) \\
&=\displaystyle\sum_{s\in V} \mathcal{S}_{V\setminus S}(s) - \mathcal{S}_V(s) \\
&=\displaystyle\sum_{s\in V} \frac{\rho(W(s))}{\rho(W)} - \frac{\rho(W(s\cup S))}{\rho(W(S))}\\
&\leq \frac{1}{\rho(W)} \displaystyle\sum_{s\in V} \rho(W(s)) - \rho(W(s\cup S))\\
&\lesssim \displaystyle\sum_{s\in V} \left[ \sqrt[n(s)]{v\left(\mathcal{C}^{n(s)}\right)} - \sqrt[n(s\cup S)]{v\left(\mathcal{C}^{n(s\cup S)}\right)} \right]
\end{align*}

\noindent where
\begin{itemize}
\item $\mathcal{S}_U(\cdot)$ is the spectral influence on the induced subgraph in $G$ by $U\subseteq V$.

\item $\mathcal{S}_U(s)=0$ for $s\not\in U$.

\item $\mathcal{C}^{m}$ is the graph dominant cycle of length $m$ and $\sqrt[m]{v\left(\mathcal{C}^{m}\right)}$ its geometric value. 
\end{itemize}

We used theorem \ref{Circuitstheorem} to establish the upper bound. Thus, $\mathbf{S}(V\setminus S)-\mathbf{S}(V)$ is the sum of each vertex's improvement/deterioration in its spectral influence after the suppression of $S$. It is bounded by the sum of the differences in the dominant cycle geometric values between the scenarios of the existence and absence of $S$, up to some constant. We have the following general result:

\begin{theorem}{\ \label{Davis-KahanEquivalent} }\\
Let $G=(V,A,W)$ be a directed graph where the vertices set $V$ has cardinal $n$, $W$ is a non negative weight matrix and $\widetilde{W}=W+Q\geq0$, for a perturbation matrix $Q$. Define the spectral cyclicality gap
\begin{align}
\label{eqn:8}
\mathbf{S}(\widetilde{W})-\mathbf{S}(W)&=\displaystyle \sum_{s\in V} \mathcal{S}_{\widetilde{W}}(s) - \mathcal{S}_{W}(s)
\end{align}
\noindent where $\mathcal{S}_Q(\cdot)$ is the spectral influence with respect to the weight matrix $Q$. Suppose $\rho(W)\geq \rho(\widetilde{W})$ (otherwise, one can interchange the rules of $\mathbf{S}(\widetilde{W})$ and $\mathbf{S}(W)$), one has
\begin{align}
\label{eqn:9}
\frac{n}{\rho(W)}\left( \left(\rho(\widetilde{W}) - \rho(W)\right) + \min_{s\in V} \left(\rho(W(s)) - \rho(\widetilde{W}(s))\right) \right)&\leq \mathbf{S}(\widetilde{W})-\mathbf{S}(W)\leq \frac{n}{\rho(W)} \max_{s\in V} \left(\rho(W(s)) - \rho(\widetilde{W}(s))\right)
\end{align}
\end{theorem}

\begin{proof}{\ }\\
Suppose $\rho(W)\geq \rho(\widetilde{W})$. On one hand
\begin{align*}
\rho(W)\left(\mathbf{S}(\widetilde{W})-\mathbf{S}(W)\right)&\geq \rho(\widetilde{W})\mathbf{S}(\widetilde{W})-\rho(W)\mathbf{S}(W)= n\left(\rho(\widetilde{W}) - \rho(W)\right)+\displaystyle\sum_{s\in V} \rho(W(s)) - \rho(\widetilde{W}(s))\\
&\geq n \left( \left(\rho(\widetilde{W}) - \rho(W)\right) + \min_{s\in V} \left(\rho(W(s)) - \rho(\widetilde{W}(s))\right) \right)\\
&\geq n \left(\rho(\widetilde{W}) - \rho(W)\right) 
\end{align*}

\noindent On the other hand
\begin{align*}
\mathbf{S}(\widetilde{W})-\mathbf{S}(W)=\displaystyle\sum_{s\in V} \frac{\rho(W(s))}{\rho(W)} - \frac{\rho(\widetilde{W}(s))}{\rho(\widetilde{W})}\leq \frac{n}{\rho(W)}\max_{s\in V} \left(\rho(W(s)) - \rho(\widetilde{W}(s))\right)
\end{align*}
\end{proof}

In the case of vertices elimination, one can use theorem \ref{Perron-Frobenius} and replace $\mathbf{S}(\widetilde{W})$ by $\mathbf{S}(V\setminus S)$ in theorem \ref{Davis-KahanEquivalent} to get the following result:

\begin{corollary}{\ }\\
Let $G=(V,A,W)$ be a weighted directed graph. One has
\begin{align}
\label{eqn:10}
\left| \mathbf{S}(V\setminus S)-\mathbf{S}(V)\right| &\leq n \max_{s\in V \cup \{\emptyset\}} \left|\mathcal{S}_{V\setminus s}(S)\right|
\end{align}
\end{corollary}

\begin{figure}[!htb]
		\begin{subfigure}{0.3\textwidth}
		\includegraphics[width=\linewidth]{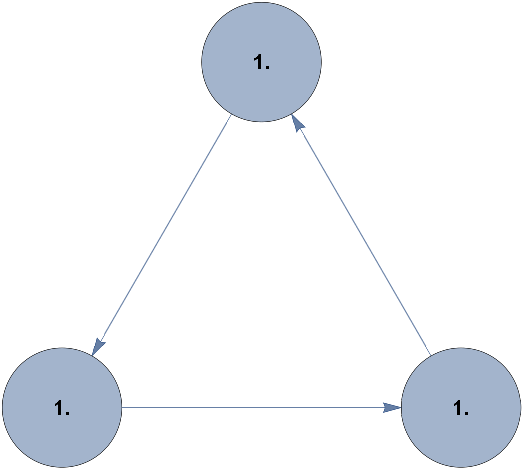}
		\caption{$\mathbf{S}=3$.}
		\label{cycle}
		\end{subfigure}\hfill
		\begin{subfigure}{0.3\textwidth}
		\includegraphics[width=\linewidth]{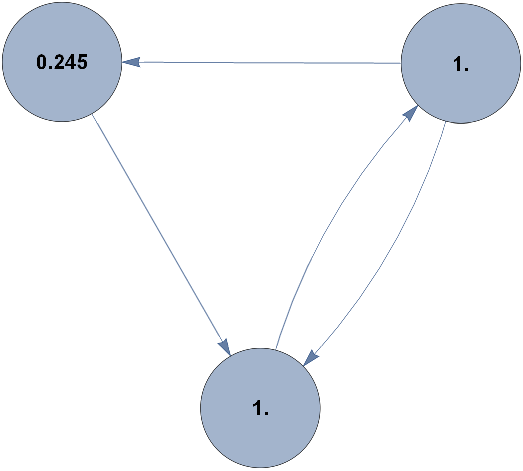}
		\caption{$\mathbf{S}=2.25$.}
		\label{circuit}
		\end{subfigure}\hfill
		\begin{subfigure}{0.3\textwidth}
		\includegraphics[width=\linewidth]{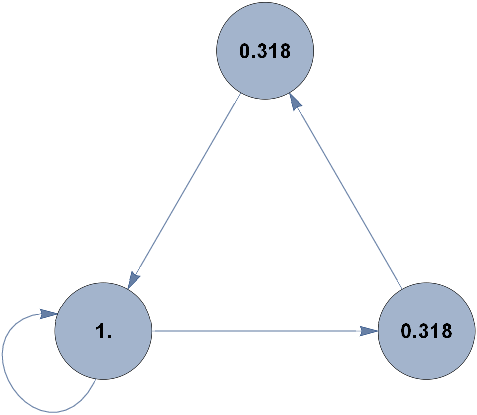}
		\caption{$\mathbf{S}=1.64$.}
		\label{loop}
		\end{subfigure}
		\vfill
		\begin{subfigure}{0.3\textwidth}
		\includegraphics[width=\linewidth]{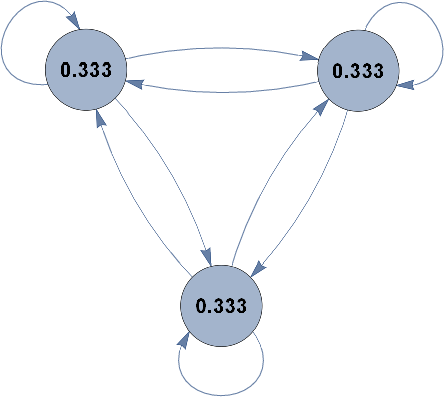}
		\caption{$\mathbf{S}=1$.}
		\end{subfigure}\hfill
		\begin{subfigure}{0.3\textwidth}
		\includegraphics[width=1.1\linewidth]{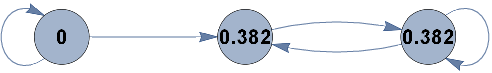}
		\caption{$\mathbf{S}=0.76$.}
		\end{subfigure}\hfill
		\begin{subfigure}{0.3\textwidth}
		\includegraphics[width=1.1\linewidth]{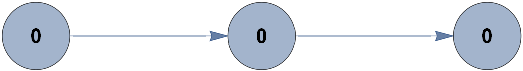}
		\caption{$\mathbf{S}=0$.}
		\end{subfigure}
		\caption{The spectral cyclicality for different dispositions of a three vertices unweighted and connected digraph: \textbf{a)} perfect diffusion on a cycle, \textbf{b)} diffusion slowed down by a circuit, \textbf{c)} diffusion slowed down by a loop, \textbf{d)} slowest diffusion in a complete graph, \textbf{e)} partial diffusion, \textbf{f)} one way diffusion.\label{Fig2}}
	\end{figure}

We conclude this section by plotting in figure \ref{Fig2} a bestiary of unweighted digraphs with the corresponding spectral influences and cyclicality.

\section{Spectral influence and cyclicality clustering\label{Spectral influence and cyclicality clustering}}

\hskip 0.5cm The spectral cyclicality is a measurement of cyclicality and diffusion on graphs. In the sequel, a clustering algorithm is defined based on this measurement to regroup vertices with strong cyclicality. To implement this algorithm, the following convention regarding isolated vertices should be adopted:

\begin{assumption}{\ }\\ \label{Singleton}
If the digraph $G=(V,A,W)$ is reduced to a unique vertex $s$, then the spectral cyclicality is $\mathbf{S}=1$.
\end{assumption}

The idea behind the assumption \ref{Singleton} is to decide whether two vertices should be split into two isolated vertices. The split decision will be made if the spectral cyclicality is not greater than $1$. Let's analyze the implication of this assumption: Given a graph with two vertices and the corresponding non-negative weight matrix \mbox{$
W=\left(
\begin{matrix}
a & b \\
c & d
\end{matrix}
\right)
$}, the spectrum of the matrix $W$ is 
$$\{\lambda_1,\lambda_2\}=\displaystyle\left\{\frac{1}{2} \left(a+d+\sqrt{(a-d)^2+4 bc}\right),\frac{1}{2} \left(a+d-\sqrt{(a-d)^2+4 bc}\right)\right\}$$

The spectral cyclicality is therefore

\begin{align*}
\mathbf{S}&=\displaystyle\frac{\frac{1}{2} \left(a+d+\sqrt{(a-d)^2+4 bc}\right)-d}{\frac{1}{2} \left(a+d+\sqrt{(a-d)^2+4 bc}\right)}+\frac{\frac{1}{2} \left(a+d+\sqrt{(a-d)^2+4 bc}\right)-a}{\frac{1}{2} \left(a+d+\sqrt{(a-d)^2+4 bc}\right)}\\
&=\displaystyle \frac{2}{\displaystyle 1+\frac{a+d}{\sqrt{(a-d)^2+4 bc}}}\\
\end{align*}

$\mathbf{S}$ reaches its maximum $2$ in the case of a cycle ($a=d=0$). If either $c$ or $b$ is zero, then $\mathbf{S}< 1$ and our algorithm split the vertices since there is no diffusion on the graph. In the general case, the splitting condition is $\frac{a+d}{\sqrt{(a-d)^2+4 bc}}\geq1$ which is equivalent to $ad\geq cd$, and this occurs when the product of loop values is greater than the cycle value.

\begin{notation}{\ }\\
We will designate by
\begin{itemize}
\item $[1,n]$ the set of integers from $1$ to $n$.
\item $\tilde{W}(S)$ the resulting matrix obtained by deleting all rows and columns of $W$, except for those that correspond to the elements of $S$.
\item $\overline{W}(S)$ the matrix resulting from $W$ by setting $w_{ij}=0$ if $i\neq j$ and $\{i,j\}\subset S$.
\end{itemize}
\end{notation}

\vskip 1cm

\subsection{Divisive cyclicality clustering}

\hskip 0.5cm We explain here how the \textbf{divisive cyclicality clustering} algorithm works: given a weighted digraph $G=(V,A,W)$, the algorithm eliminates the vertex whose removal increases spectral cyclicality the most. If no increase is possible, the algorithm eliminates the vertices whose removal does not decrease spectral cyclicality. The same operation is applied to deleted vertices until all vertices are classified.

\vskip 1cm

\begin{tabular}{cc}
\hline
\textbf{Algorithm 1:} & Divisive cyclicality clustering \\
\hline
\textbf{Input:} & a weight matrix $W\in M_n$ and a vertices set $V$\\
\textbf{Initialization:} & set $S=\{\}$\\
\textbf{While:} & $\displaystyle\max_{i\in V}\mathbf{S}(W(i)) \geq \mathbf{S}(W)$ \\
  & update $W=W(i^{\star})$ , $S= S \cup \{i^{\star}\}$,  where $i^{\star}=\displaystyle\arg\max_{i\in V}\mathbf{S}(W(i))$\\
\textbf{Update:} & set $V=S$.\\
\textbf{Output:} & a partition $\{V_1,\hdots,V_k\}$.\\
\end{tabular}

\subsection{Agglomerative cyclicality clustering}

\hskip 0.5cm Given a digraph $G=(V,A,W)$, the \textbf{agglomerative cyclicality clustering} algorithm groups vertices based on their spectral cyclicality, with the highest values being grouped first. If no increase in cyclicality is possible, the algorithm repeats the process with the remaining vertices until all vertices are classified.

\vskip 1cm

\begin{tabular}{cc}
\hline
\textbf{Algorithm 2:} & Agglomerative cyclicality clustering \\
\hline
\textbf{Input:} & a weight matrix $W\in M_n$ and a vertices set $V$\\
\textbf{Initialization:} & set $S=\displaystyle\arg\max_{(j,k)\in V^2}\mathbf{S}(\tilde{W}(\{j,k\}))$\\
\textbf{While:} & $\displaystyle \max_{i\in V}\mathbf{S}(\tilde{W}(S\cup\{i\})) > \mathbf{S}(\tilde{W}(S))$ \\
  & update $S=S\cup\{i^\star\}$, where $i^{\star}=\displaystyle\arg\max_{i\in V}\mathbf{S}(\tilde{W}(S\cup\{i\}))$\\
\textbf{Update:} & set $V=V\setminus  S$.\\
\textbf{Output:} & a partition $\{V_1,\hdots,V_k\}$.\\
\end{tabular}

\subsection{Overlapping clustering}

\hskip 0.5cm Let's denote by $S_1(W)$ the first partition obtained by applying the divisive (resp. agglomerative) cyclicality algorithm. Overlapping clustering can be achieved by preserving the vertices and deleting the partition intra-partition connections at each stage:

\vskip 1cm

\begin{tabular}{cc}
\hline
\textbf{Algorithm 3:} & Overlapping divisive/agglomerative cyclicality clustering \\
\hline
\textbf{Input:} & a weight matrix $W\in M_n$ and a vertices set $V$\\
\textbf{Initialization:} & set $O=S_1(W)$\\
\textbf{While:} & $\displaystyle \mathbf{S}(\overline{W}(O))>1$ \\
  & update $W=\overline{W}(O)$ , $O=S_1(W)$\\
\textbf{Output:} & an overlapping partition $\{O_1,\hdots,O_k\}$.\\
\end{tabular}

\subsection{Discussion}

\hskip 0.5cm Divisive and agglomerative clustering schemes are fundamentally different in nature. Divisive clustering starts with a single cluster that contains all the vertices, and then recursively splits it into smaller clusters, whereas agglomerative clustering starts with each vertex as a separate cluster and then merges them into larger clusters.

Overall, both divisive and agglomerative cyclicality clustering share the same philosophy of grouping vertices with high cyclicality, even though they achieve this goal through different methods. Let's analyze how the algorithm practically works in the case of a divisive scheme:

Following remark \ref{SIsituations}.5, the algorithm starts deleting vertices not strongly connected to the dominant cycle, then it starts deleting competitor cycles reducing the cyclicality. For unweighted graphs, the algorithm chooses the longest cycle as the dominant one. It is important to note that the algorithm’s outcome may not be a perfect cycle, but a strong cyclicality component, depending on the structure of the network. When overlapping is allowed, the algorithm cuts off all the arcs between the elements of a partition keeping the loops, and preventing the same cycles from appearing in other
partitions.

One limitation of the cyclicality clustering algorithm is the tie situation where two or more vertices have an equal effect on improving cyclicality. The algorithm picks a vertex randomly; moreover, the clustering process depends on the division/agglomeration path, i.e., the order of vertex suppression/concatenation. Consequently, Some vertices may be eliminated or regrouped too early, resulting in a component that is not cyclically optimal.

\subsection{Benchmark}

\hskip 0.5cm The spectral cyclicality clustering algorithm should not be confused with the spectral clustering technique. While both techniques utilize eigengaps, their purposes are different: spectral clustering uses the eigenvectors of the Laplacian matrix to reduce the dimensionality of the data, whereas eigengaps, which are the differences between consecutive eigenvalues, are used as a heuristic to determine the number $k$ of partitions, allowing a more effective application of connectivity-based clustering algorithms like $k$-means. This contrasts with our technique, where eigengaps, defined as the difference between the first eigenvalues for different situations, are used to measure the importance of a vertex in diffusion within a network, with diffusion, rather than connectivity, being the primary objective.

In the sequel, we compare our clustering algorithm with traditional community detection algorithms. Specifically, we will compare our divisive cyclicality clustering algorithm (CYC) with three distinct algorithms. For a comprehensive overview of these methods, we refer the reader to Fortunato's article \cite{Fortunato2010}.

\begin{enumerate}
\item \textbf{Centrality clustering} (CC): The algorithm involves removing edges from a graph based on their betweenness centrality values, which measure the number of shortest paths between all vertex pairs that pass through the edge. This approach highlights edges that are crucial for connectivity and flow within the network.

Initially, the centrality for all edges is computed. Then, the edge with the highest centrality is removed; if there are ties, one is chosen at random. The centralities are recalculated on the modified graph, and this process is repeated.

\item \textbf{Modularity clustering} (MC): The goal is to partition the network into clusters such that the density of edges within communities is higher than the density of edges between communities. The concept of modularity is central as it provides a quantitative measure of the strength of the division of a network into communities.

\item \textbf{Spectral clustering} (SC): Spectral clustering involves transforming the initial set of objects into a new set of points in a space where the coordinates are derived from the eigenvectors. These transformed points are then grouped using the k-means method.
\end{enumerate}

We begin the benchmark by comparing our algorithms on five classical unweighted directed graphs. The clustering results are depicted in table \ref{Tab1}.

\begin{center}
\begin{table}[!htb]
\setlength\extrarowheight{15pt}
\begin{tabular}{|l|*{4}{c|}}
\hline
\backslashbox{\textbf{Object}}{\textbf{Method}} & \textbf{CYC} & \textbf{CC} & \textbf{MC} & \textbf{SC} \\\hline\hline
\includegraphics[width=0.18\linewidth]{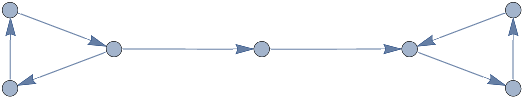} &
\includegraphics[width=0.18\linewidth]{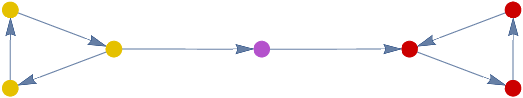} & \includegraphics[width=0.18\linewidth]{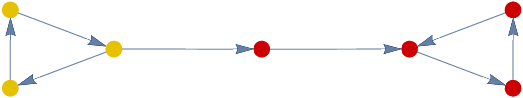} & \includegraphics[width=0.18\linewidth]{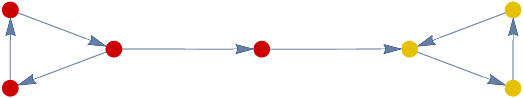} & \includegraphics[width=0.18\linewidth]{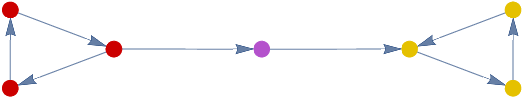} \\[5pt]
\hline
\includegraphics[width=0.15\linewidth]{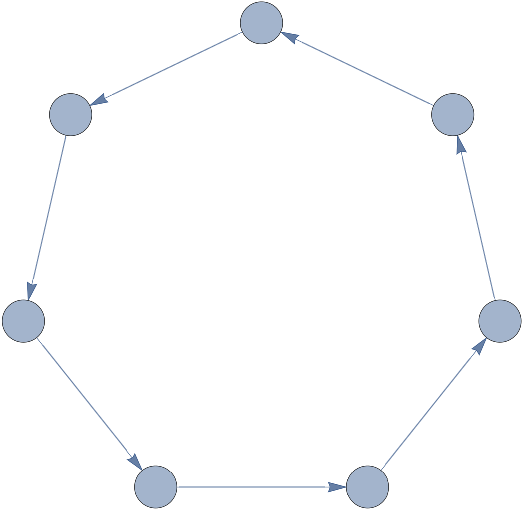} &
\includegraphics[width=0.15\linewidth]{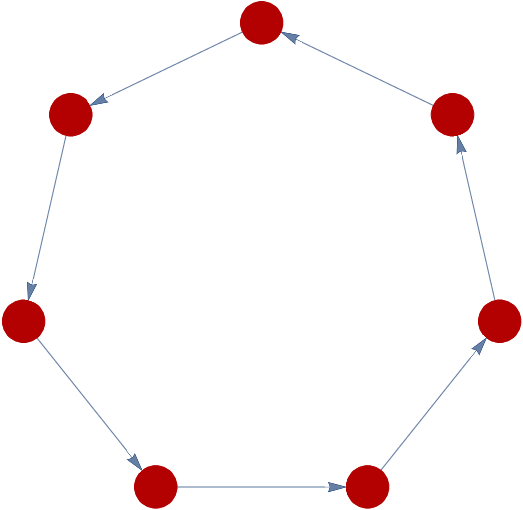} & \includegraphics[width=0.15\linewidth]{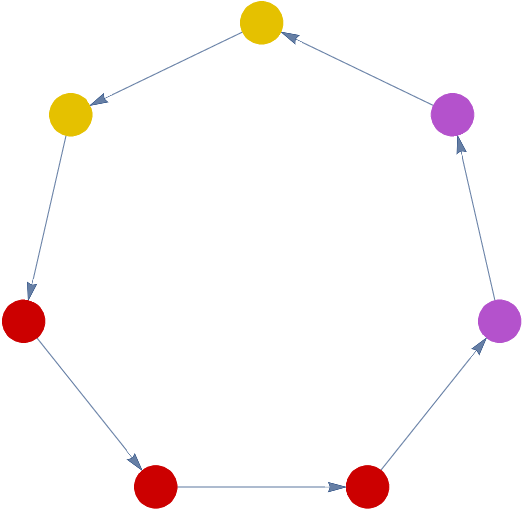} & \includegraphics[width=0.15\linewidth]{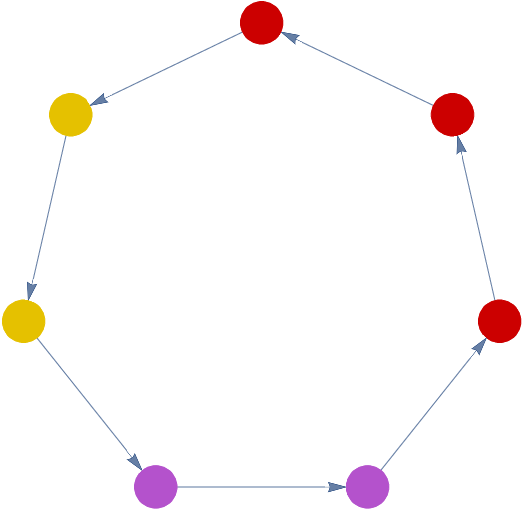} & \includegraphics[width=0.15\linewidth]{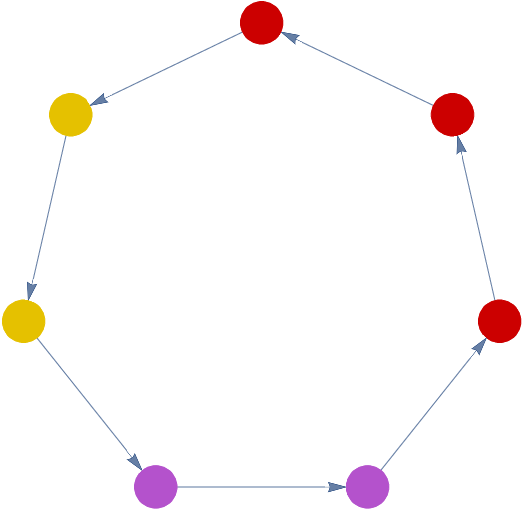} \\
\hline
\includegraphics[width=0.15\linewidth]{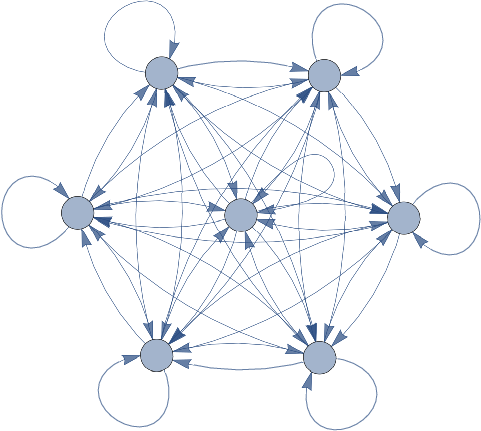} &
\includegraphics[width=0.15\linewidth]{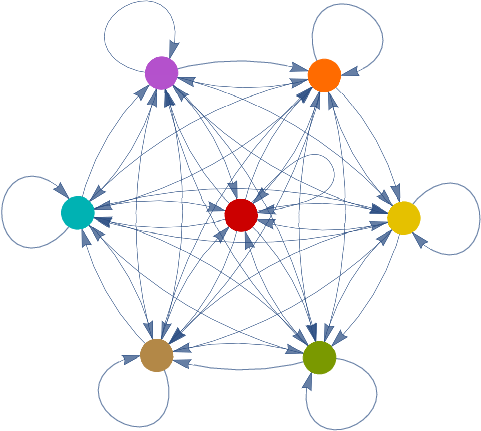} & \includegraphics[width=0.15\linewidth]{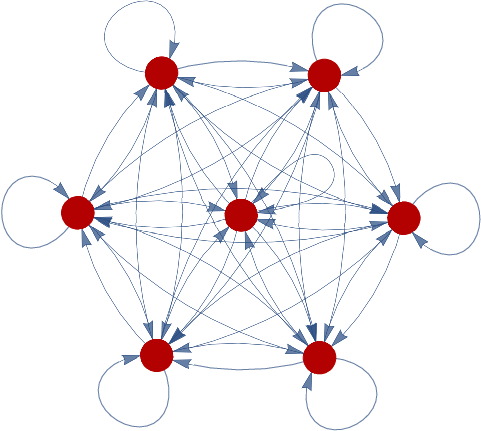} & \includegraphics[width=0.15\linewidth]{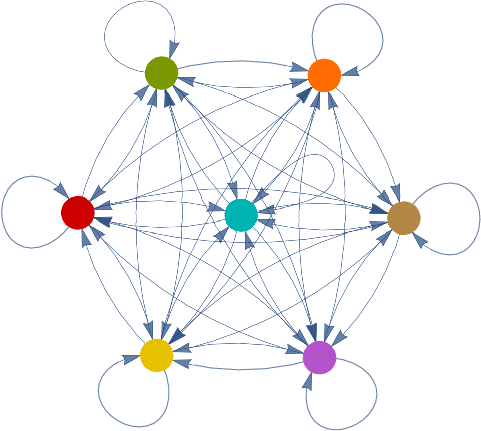} & \includegraphics[width=0.15\linewidth]{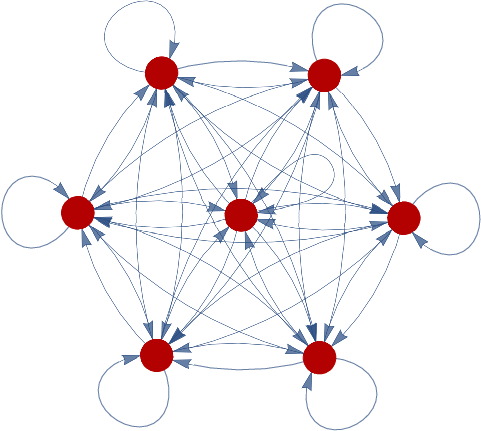} \\
\hline
\includegraphics[width=0.15\linewidth]{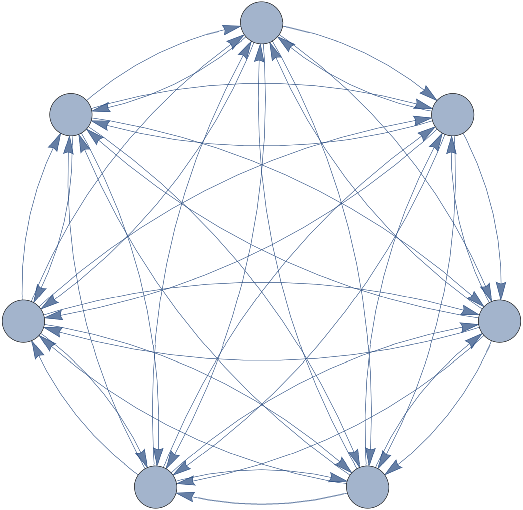} &
\includegraphics[width=0.15\linewidth]{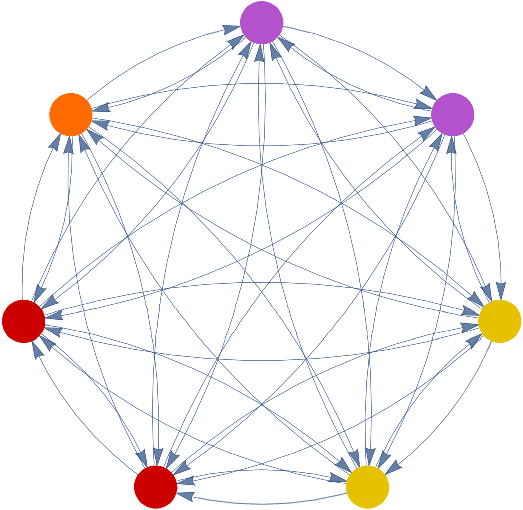} & \includegraphics[width=0.15\linewidth]{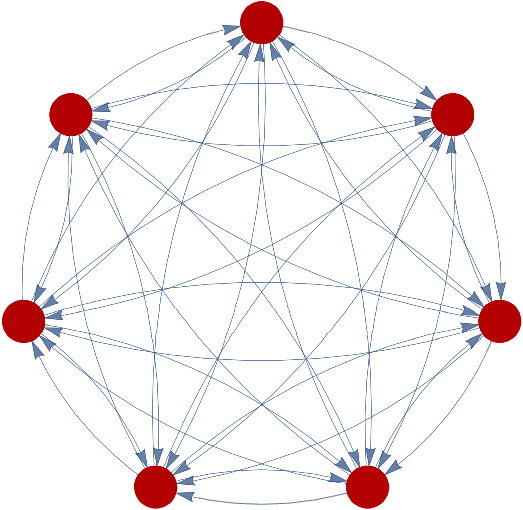} & \includegraphics[width=0.15\linewidth]{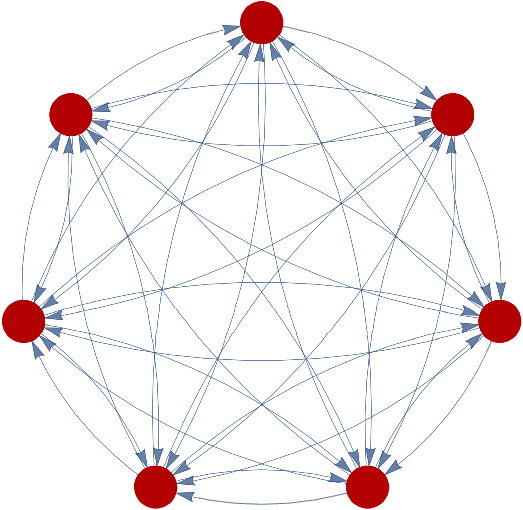} & \includegraphics[width=0.15\linewidth]{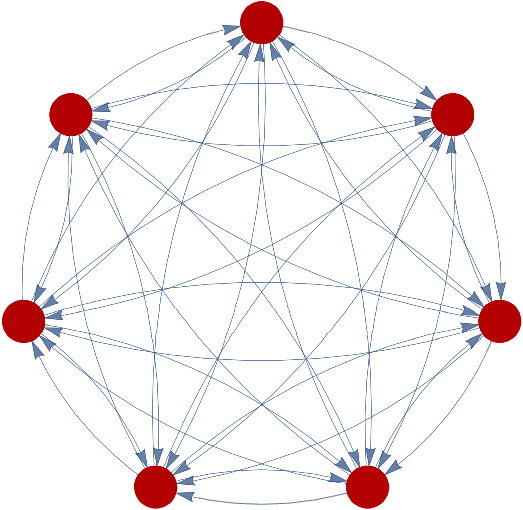} \\
\hline
\includegraphics[width=0.18\linewidth]{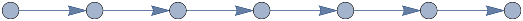} &
\includegraphics[width=0.18\linewidth]{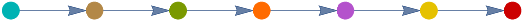} & \includegraphics[width=0.18\linewidth]{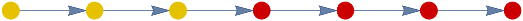} & \includegraphics[width=0.18\linewidth]{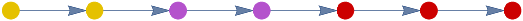} & \includegraphics[width=0.18\linewidth]{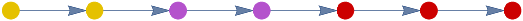} \\[10pt]
\hline
\end{tabular}

\caption{Resulting clustering of typical directed graphs using four clustering algorithms: spectral cyclicality (CYC), centrality (CC), modularity (MC) and spectral (SC).\label{Tab1}}
\end{table}
\end{center}

The reader should be able to distinguish the spectral cyclicality clustering algorithm from the other techniques. The most noticeable differences occur in objects two and four, where the other algorithms divide the cycle and keep the complete graph compact. This distinction arises from differing clustering paradigms: while our algorithm prioritizes \textbf{diffusion}, the other algorithms emphasize \textbf{connectivity}. We provide commentary on our technique for each object:

\begin{itemize}
\item \textbf{Object 1}: The spectral cyclicality clustering divides the first graph into two diffusion components (cycles) and one vertex isolated from diffusion.

\item \textbf{Object 2}: The algorithm preserves the cycle, which represents the essence of our clustering technique.

\item \textbf{Object 3}: A complete graph with loops is entirely divided by our algorithm, as the multitude of circuits slows down diffusion. In this case, loops are the most effective diffusion objects.

\item \textbf{Object 4}: Without loops, a complete graph is divided into binary cycles, with one randomly isolated vertex.

\item \textbf{Object 5}: A chain is fully divided by our algorithm due to the lack of diffusion.

\end{itemize}

To perform benchmarking, we propose applying clustering techniques to a large sample of networks. The networks are unweighted, directed graphs with $n\times n$ adjacency matrices generated randomly. We will use three cluster quality metrics to evaluate the algorithms:

\begin{enumerate}
\item \textbf{Cyclicality}: We define cyclicality to be the average
\begin{align}
\label{eqn:10}
\mathbb{S} &= \frac{\sum_{l=1}^k  \mathbf{S}(V_l)}{k}
\end{align}

where:
\begin{itemize}
  \item $\mathbf{S}(V_l)$ is the spectral cyclicality of community $V_l$, as introduced in definition \ref{spectral cyclicality}.
  \item \( k \) is the number of communities.
\end{itemize}

\item \textbf{Modularity}: The modularity of a graph compares the presence of each intra-cluster edge of the graph with the probability that that edge would exist in a random graph. It is defined by the formula:
\begin{align}
\label{eqn:11}
\mathbb{Q} &= \frac{1}{m} \sum_{i,j} \left( W_{ij} - \frac{k_i^{\text{out}} k_j^{\text{in}}}{m} \right) \delta(V_i, V_j)
\end{align}

where:
\begin{itemize}
  \item \( W_{ij} \) is the weight of the arc from vertex \( i \) to vertex \( j \) in the weighted adjacency matrix.
  \item \( k_i^{\text{out}} \) is the weighted out-degree of vertex \( i \), calculated as \( \sum_j W_{ij} \).
  \item \( k_j^{\text{in}} \) is the weighted in-degree of vertex \( j \), calculated as \( \sum_i W_{ij} \).
  \item \( m \) is the total weight of all arcs in the graph, calculated as \( m = \sum_{i,j} W_{ij} \).
  \item \( \delta(V_i, V_j) \) is the Kronecker delta, which is 1 if vertices \( i \) and \( j \) are in the same community, and 0 otherwise.
\end{itemize}

\item \textbf{Coverage}: The coverage compares the fraction of intra-cluster edges in the graph to the total number of edges in the graph. It is given by
\begin{align}
\label{eqn:12}
\mathbb{C} &= \frac{\sum_{l=1}^k \sum_{i, j \in V_l} W_{ij}}{\sum_{i, j} W_{ij}}
\end{align}

where:
\begin{itemize}
  \item \( k \) is the number of communities.
  \item \(W_{ij} \) is the weight of the arc from vertex \( i \) to vertex \( j \).
  \item \( \sum_{l=1}^k \sum_{i, j \in V_l} W_{ij} \) represents the total weight of arcs within all communities.
  \item \( \sum_{i, j} W_{ij} \) is the total weight of all arcs in the graph.
\end{itemize}
\end{enumerate}

We generated $100$ networks for each size $n = 5, 10, 15, 20$ and $25$. The results of our simulation are summarized in the table below:

\begin{center}
\begin{table}[!htb]
\begin{tabular}{|l|*{3}{c|}}
\hline
\backslashbox{\textbf{Dimension}}{\textbf{Metric}} & \textbf{Cyclicality} & \textbf{Modularity} & \textbf{Coverage} \\\hline\hline
\rule{0pt}{16ex} $n=5$ &
\includegraphics[width=0.23\linewidth]{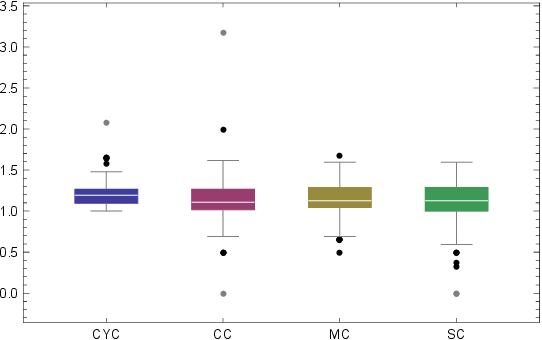} & \includegraphics[width=0.23\linewidth]{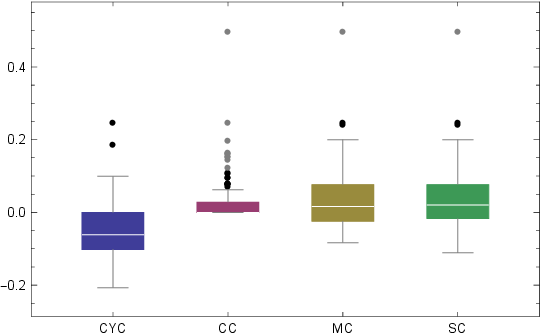} & \includegraphics[width=0.23\linewidth]{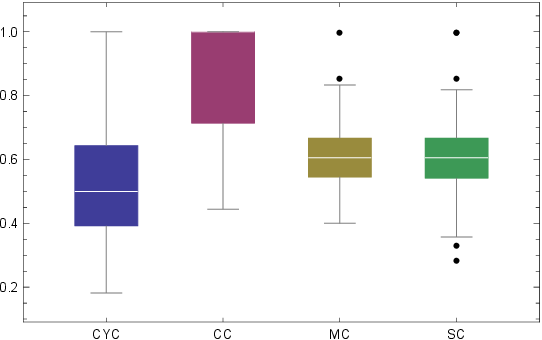} \\
\hline
\rule{0pt}{16ex} $n=10$ &
\includegraphics[width=0.23\linewidth]{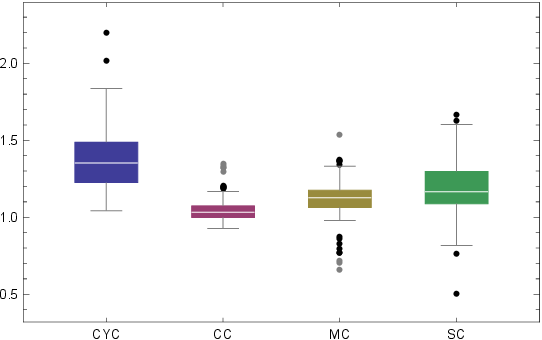} & \includegraphics[width=0.23\linewidth]{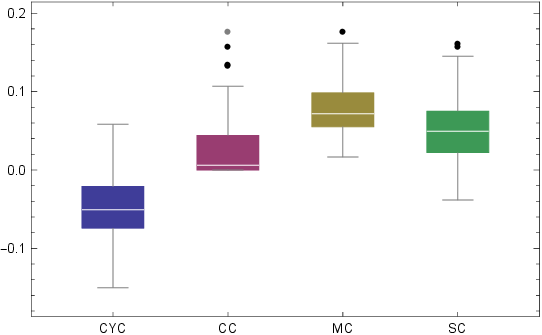} & \includegraphics[width=0.23\linewidth]{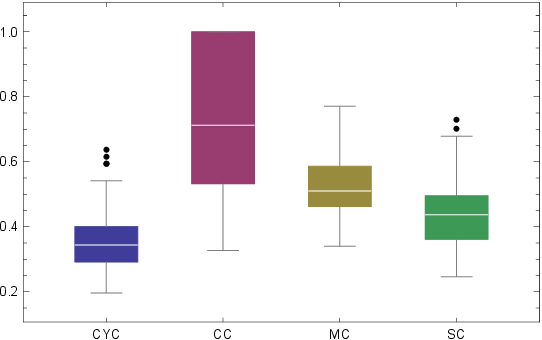} \\
\hline
\rule{0pt}{16ex} $n=15$ &
\includegraphics[width=0.23\linewidth]{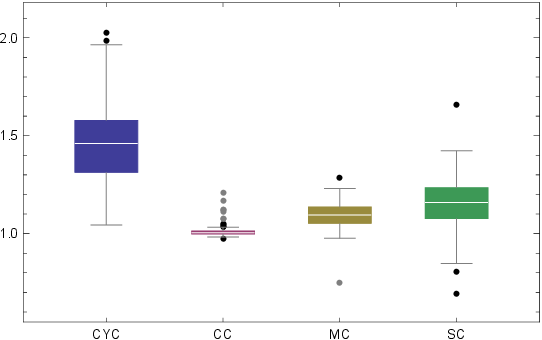} & \includegraphics[width=0.23\linewidth]{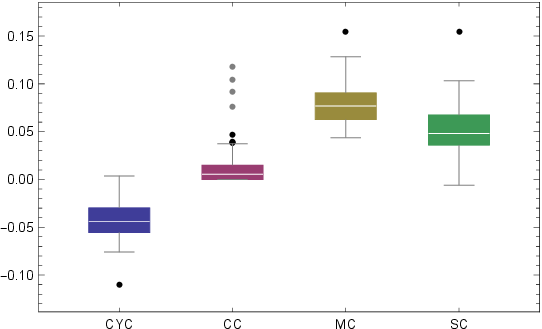} & \includegraphics[width=0.23\linewidth]{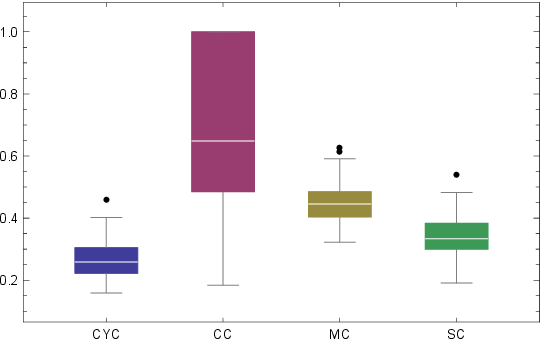} \\
\hline
\rule{0pt}{16ex} $n=20$ &
\includegraphics[width=0.23\linewidth]{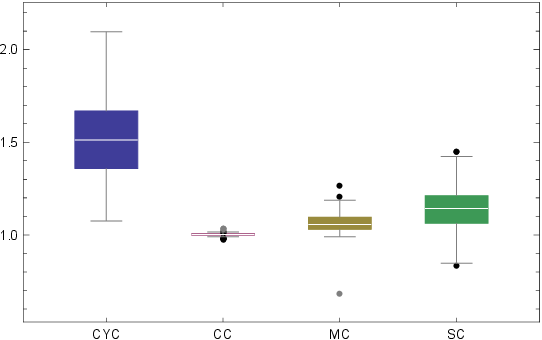} & \includegraphics[width=0.23\linewidth]{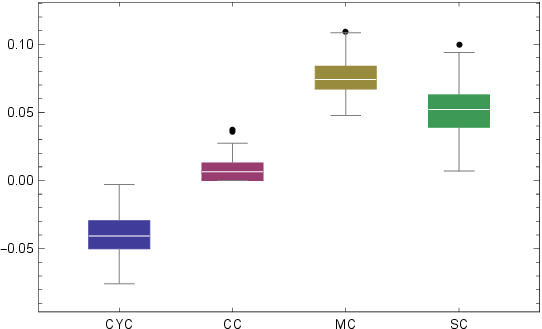} & \includegraphics[width=0.23\linewidth]{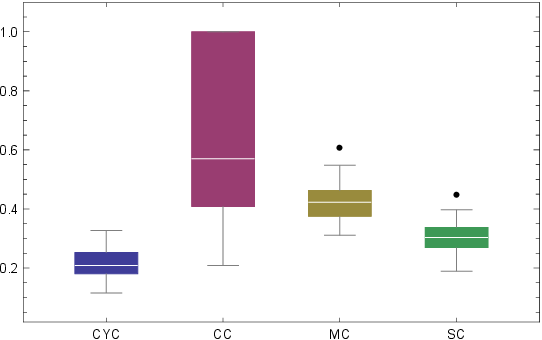} \\
\hline
\rule{0pt}{16ex} $n=25$ &
\includegraphics[width=0.23\linewidth]{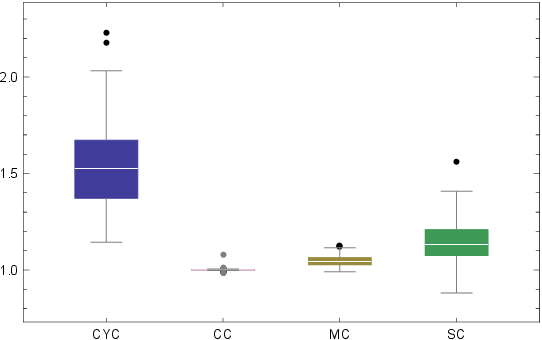} & \includegraphics[width=0.23\linewidth]{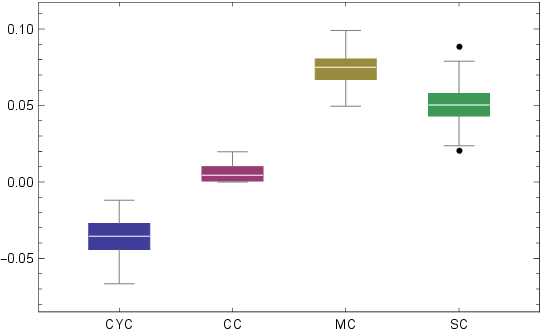} & \includegraphics[width=0.23\linewidth]{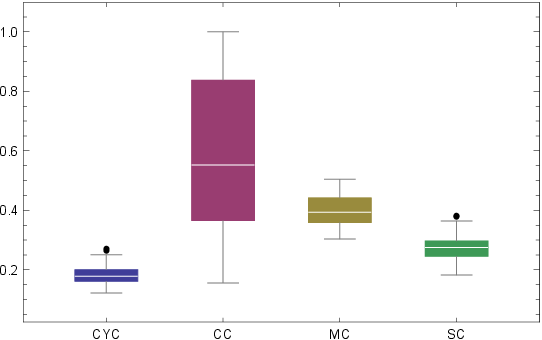} \\
\hline
\end{tabular}
\caption{Boxplot of the simulation results for graphs with different values of $n$.\label{Tab2}}
\end{table}
\end{center}

As expected, our algorithm is the most efficient in terms of cyclicality, followed by the spectral clustering technique. As the dimension $n$ increases, our algorithm seems to outperform the others but starts to lose accuracy. Similarly, modularity clustering demonstrates the best performance in terms of modularity. Centrality clustering, while improving betweenness centrality, tends to reach a centrality value of $1$ (complete graph). The spectral cyclicality algorithm performs poorly with respect to modularity and coverage criteria. As explained earlier, this is due to the difference between the cyclicality paradigm and the connectivity paradigm. This is evident in the coverage criterion, which shows that our algorithm prioritizes diffusion over interconnectivity.

One final consideration pertains to the clustering of weighted graphs. The algorithm's results can vary significantly depending on the weight distribution. We will delve into this situation in the next section.

\section{Application to input-output analysis\label{Application to input-output analysis}}

\hskip 0.5cm Applying spectral cyclicality to input-output analysis helps us better understand interindustrial flows and connections between industrial poles. On the one hand, some poles plays the role of propulsive industries driving other poles to increase their production and creating transformation cycles, such mechanism is synonym with interdependence, influence, shock diffusion and amplification effects. On the other hand, some industries could form a complex characterized by a transformation cycle and high added-value.

Unlike input-output analysis based on Leontief or Ghosh normalization, the spectral technique does not require a direction of normalization (supply or demand) and is applied directly to the absolute flow.

In the sequel, we analyze the Moroccan interindustrial flows of the OECD Input-Output Tables (IOTs)~\cite{IOTsOCDE}, from $1995$ to $2018$. The data traces flows of intermediate goods and services of $45$ industrial poles in millions US Dollars (see appendix \ref{appendix1}).

Fix~$n=45$, the number of industrial poles, where, for~\mbox{$1 \leq i \leq n$}, the production of the~\mbox{$i^{th}$} pole is denoted by~\mbox{$X_i$}, while its final consumption is~\mbox{$Y_i$}.  The production is shared, with intermediary consumption~\mbox{$w_{ij}$} between the~\mbox{$i^{th}$} and~\mbox{$j^{th}$} poles,~\mbox{$1 \leq i,j \leq n$}. The input-output system can be written as:
\begin{align}
\label{eqn:13}
\forall \, (i,j)\, \in\, \left \lbrace 1,\hdots,n \right \rbrace^2\, :\quad X_i- \sum_{j=1}^n w_{ij}  = Y_i
\end{align}

We denote by~\mbox{$W=(w_{ij})_{1\leq i,j\leq n}$} the intermediary consumption matrix, where~\mbox{$w_{ij}\geq 0$} stands for the outflow from pole~$i$ to pole~$j$. $W$ can be understood as a weight matrix corresponding to a weighted directed graph~\mbox{$G=(V,A,W)$}, where $V$ denotes the set of industrial poles, and~$A$ is the set of arcs (connections).

We refer to a \textbf{propulsive industry} as an industrial pole with high spectral influence. Such a pole exerts a strong diffusion effect and plays a central role in the production cycle.

We refer to a \textbf{propulsive cluster} as a group of industrial poles with high spectral cyclicality, indicating strong interdependence and high added value.

\subsection{Moroccan interindustrial cyclicality analysis}

\begin{figure}[!htb]
	\begin{center}
	\includegraphics[width=0.6\linewidth]{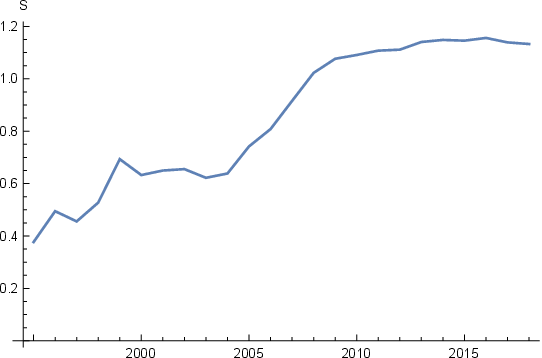}	
	\end{center}
	\caption{Evolution of Moroccan spectral cyclicality from $1995$ to $2018$.}
	\label{Fig3}
	\end{figure}

\hskip 0.5cm We present in figure \ref{Fig3} the evolution of spectral cyclicality of Moroccan interindustrial exchanges. The curve illustrates a clear upward trend in cyclicality.

A benchmark analysis is conducted for a pool of $30$ countries, each country is identified by its three-letters ISO code\footnote{International Organization for Standardization \url{https://www.iso.org/iso-3166-country-codes.html}.}. It is noteworthy that Morocco exhibits a distinct cyclical pattern:

\begin{table}[!htb]
\begin{tabular}{|c|c|c|c|c|c|c|c|c|c|c|c|c|}
\hline
 & \textbf{ARG} &  \textbf{AUS} &  \textbf{BEL} &  \textbf{BRA} &  \textbf{CAN} &  \textbf{CHN} &  \textbf{COL} &  \textbf{CZE} & \textbf{DEU} & \textbf{DNK} & \textbf{FRA} & \textbf{GBR} \tabularnewline
\hline
$\mathbf{S}$ & $0.687$ & $0.599$ & $0.363$ & $0.663$  & $0.783$ & $0.297$ & $0.538$ & $0.480$ & $0.379$ & $0.872$ & $0.566$ & $0.377$ \tabularnewline
\hline
\end{tabular}

\vskip 0.25cm

\begin{tabular}{|c|c|c|c|c|c|c|c|c|c|c|c|c|}
\hline
 & \textbf{HUN} & \textbf{ITA} & \textbf{JPN} &  \textbf{KAZ} &  \textbf{MAR} &  \textbf{MEX} &  \textbf{MYS} &  \textbf{POR} &  \textbf{RUS} &  \textbf{SAU} &  \textbf{SPA} &  \textbf{THA} \tabularnewline
\hline
$\mathbf{S}$ & $0.445$ & $0.729$ & $0.426$  & $0.387$ & $1.133$ & $0.349$ & $0.957$ & $0.293$ & $0.776$ & $1.024$ & $0.750$ & $0.398$ \tabularnewline
\hline
\end{tabular}

\vskip 0.25cm

\begin{tabular}{|c|c|c|c|c|c|c|c|c|c|c|c|c|}
\hline
 &  \textbf{TUN} &  \textbf{TUR} &  \textbf{SGP} &  \textbf{USA}  &  \textbf{VNM} & \textbf{ZAF} \tabularnewline
\hline
$\mathbf{S}$  & $0.217$ & $0.200$ & $0.331$ & $0.615$ & $0.589$ & $0.617$  \tabularnewline
\hline
\end{tabular}
\captionof{table}{Spectral cyclicality by country.}
\label{Tab3}
\end{table}

\subsection{Propulsive industries and key cyclicality components}

\hskip 0.5cm Next, we will perform a spectral cyclicality analysis of the Moroccan interindustrial flows from $1995$ to $2018$. The Tables in the appendix \ref{appendix2} reports, for each industry, the spectral influence $\mathcal{S}$ and the corresponding cyclical cluster in the case of non-overlapping divisive (\textbf{DC}) and agglomerative (\textbf{AC}) clustering, we put the cluster number of each pole, and the Greek letter $\iota$ if the pole is isolated. The results illustrate the profound changes in the interindustrial structure.

Applying cyclicality clustering algorithms, the interindustrial flows in Morocco for $2018$ could be grouped and ordered from a cyclicality perspective in one of four ways (see the appendix \ref{appendix3} for an explicit description of each component). We will visualize next the principal (non atomic) cyclicality components as a graph, with arcs represented proportionally to their weights and vertices represented based on their spectral influence within the component.

\begin{figure}[!htb]
	\begin{center}
	\begin{subfigure}{0.475\textwidth}
	\includegraphics[width=\linewidth]{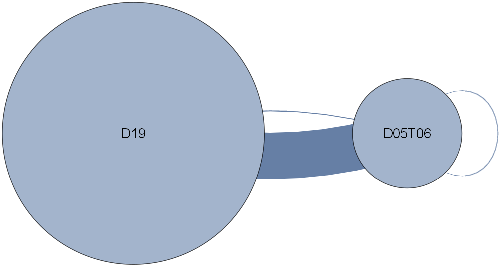}
	\caption{$\mathbf{S}=1.35676$.}
	\end{subfigure}
	\begin{subfigure}{0.475\textwidth}
	\includegraphics[width=\linewidth]{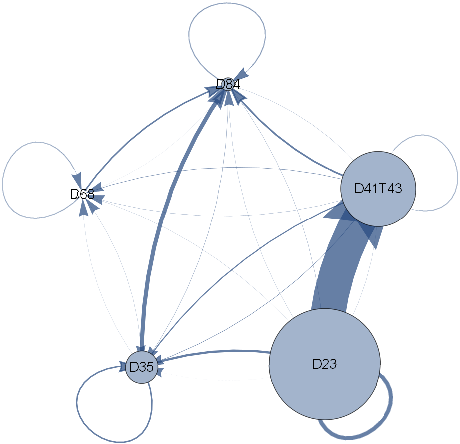}
	\caption{$\mathbf{S}=1.27431$.}
	\end{subfigure}
	\begin{subfigure}{0.475\textwidth}
	\includegraphics[width=\linewidth]{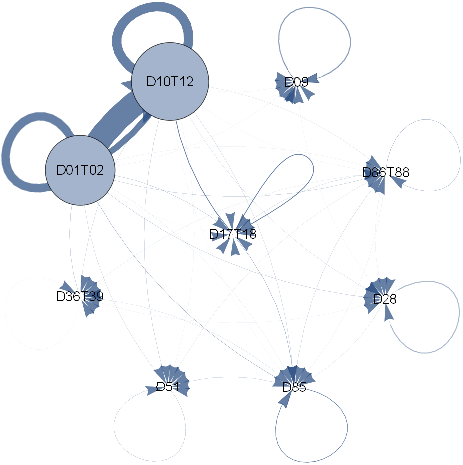}
	\caption{$\mathbf{S}=1.16459$.}
	\end{subfigure}
	\begin{subfigure}{0.475\textwidth}
	\includegraphics[width=\linewidth]{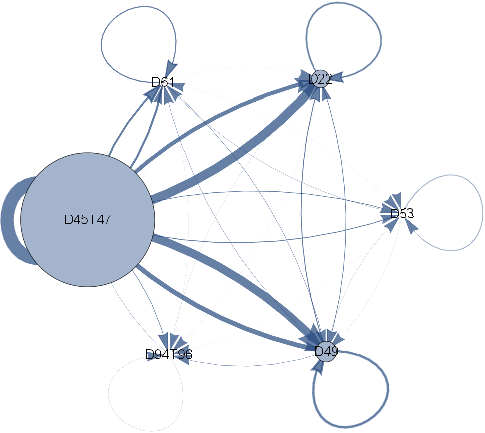}
	\caption{$\mathbf{S}=1.11335$.}
	\end{subfigure}
	\begin{subfigure}{0.475\textwidth}
	\includegraphics[width=\linewidth]{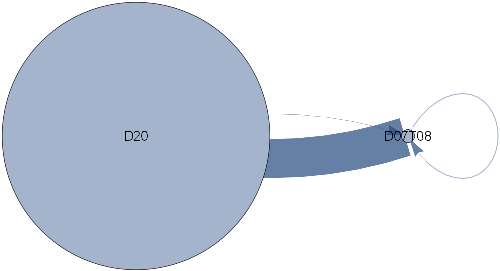}
	\caption{$\mathbf{S}=1.03187$.}
	\end{subfigure}
	\end{center}
	\caption{The five divisive clusters of $2018$ Moroccan input-output table.}
	\label{Fig4}
	\end{figure}

\begin{figure}[!htb]
	\begin{center}
	\begin{subfigure}{0.475\textwidth}
	\includegraphics[width=\linewidth]{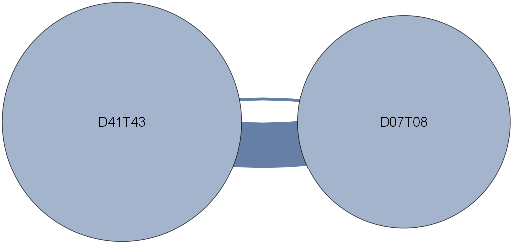}
	\caption{$\mathbf{S}=1.6033$.}
	\end{subfigure}
	\begin{subfigure}{0.475\textwidth}
	\includegraphics[width=\linewidth]{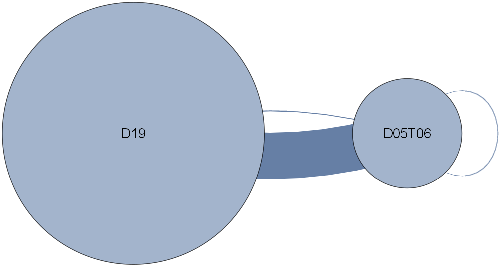}
	\caption{$\mathbf{S}=1.35676$.}
	\end{subfigure}
	\begin{subfigure}{0.475\textwidth}
	\includegraphics[width=\linewidth]{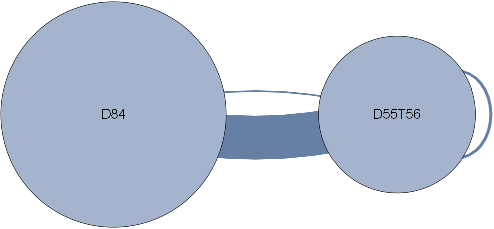}
	\caption{$\mathbf{S}=1.34825$.}
	\end{subfigure}
	\begin{subfigure}{0.475\textwidth}
	\includegraphics[width=\linewidth]{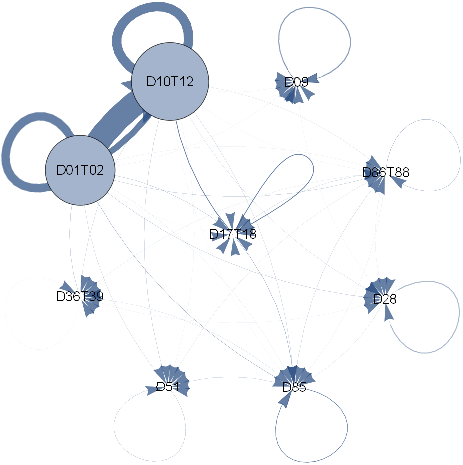}
	\caption{$\mathbf{S}=1.16459$.}
	\end{subfigure}
	\begin{subfigure}{0.475\textwidth}
	\includegraphics[width=\linewidth]{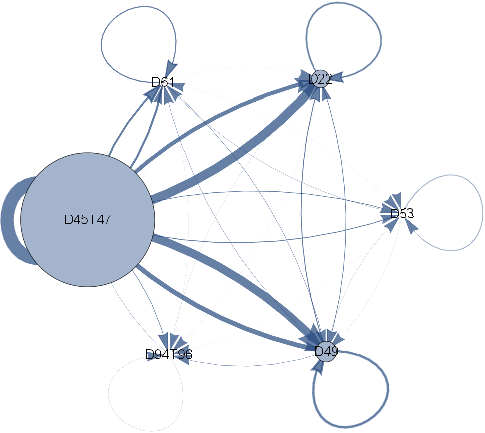}
	\caption{$\mathbf{S}=1.11335$.}
	\end{subfigure}
	\begin{subfigure}{0.475\textwidth}
	\includegraphics[width=\linewidth]{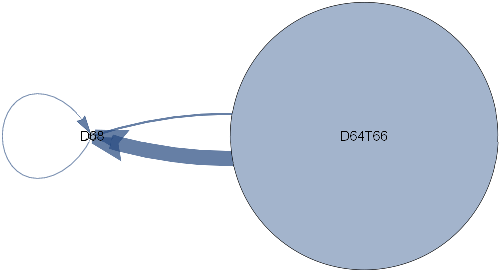}
	\caption{$\mathbf{S}=1.00877$.}
	\end{subfigure}
	\end{center}
	\caption{The six agglomerative clusters of $2018$ Moroccan input-output table.}
	\label{Fig5}
	\end{figure}

\begin{figure}[!htb]
	\begin{center}
	\begin{subfigure}{0.455\textwidth}
	\includegraphics[width=\linewidth]{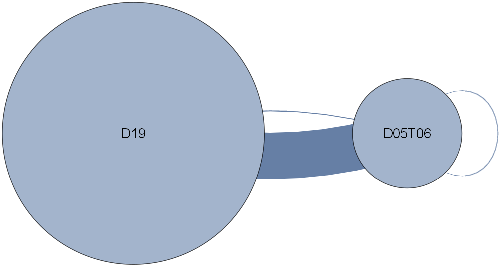}
	\caption{$\mathbf{S}=1.35676$.}
	\end{subfigure}
	\begin{subfigure}{0.455\textwidth}
	\includegraphics[width=\linewidth]{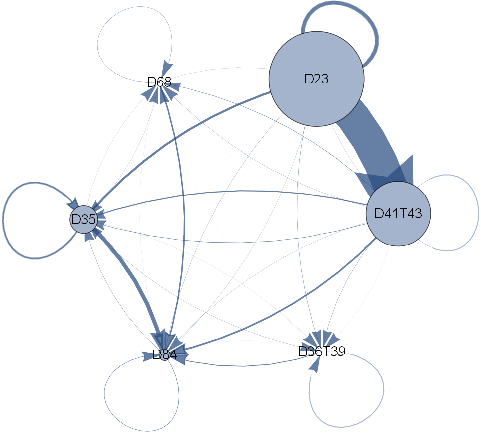}
	\caption{$\mathbf{S}=1.27467$.}
	\end{subfigure}
	\begin{subfigure}{0.455\textwidth}
	\includegraphics[width=\linewidth]{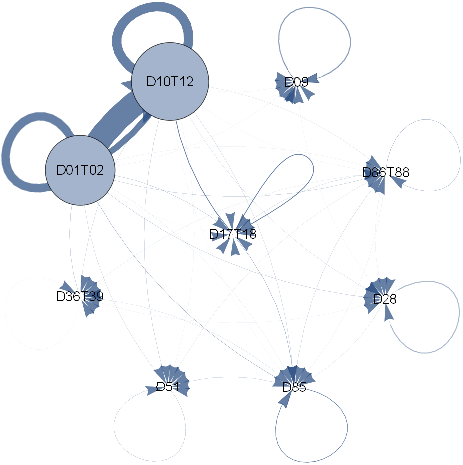}
	\caption{$\mathbf{S}=1.16459$.}
	\end{subfigure}
	\begin{subfigure}{0.455\textwidth}
	\includegraphics[width=\linewidth]{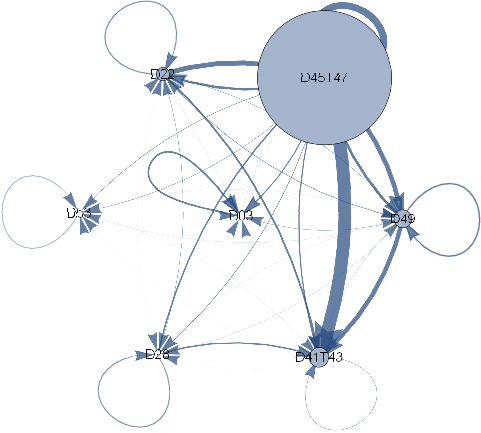}
	\caption{$\mathbf{S}=1.15275$.}
	\end{subfigure}
	\begin{subfigure}{0.455\textwidth}
	\includegraphics[width=\linewidth]{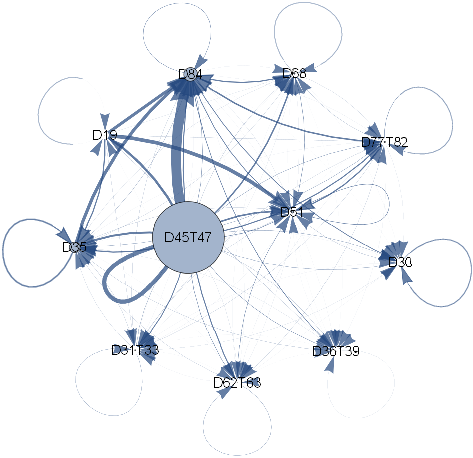}
	\caption{$\mathbf{S}=1.02185$.}
	\end{subfigure}
	\end{center}
	\caption{The five overlapping divisive clusters of $2018$ Moroccan input-output table.}
	\label{Fig6}
	\end{figure}

\begin{figure}[!htb]
	\begin{center}
	\begin{subfigure}{0.475\textwidth}
	\includegraphics[width=\linewidth]{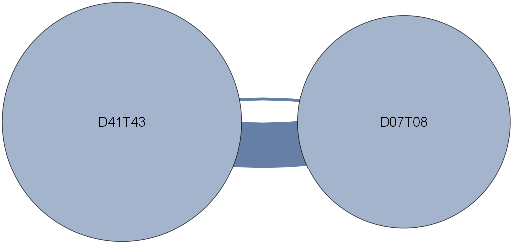}
	\caption{$\mathbf{S}=1.6033$.}
	\end{subfigure}
	\begin{subfigure}{0.475\textwidth}
	\includegraphics[width=\linewidth]{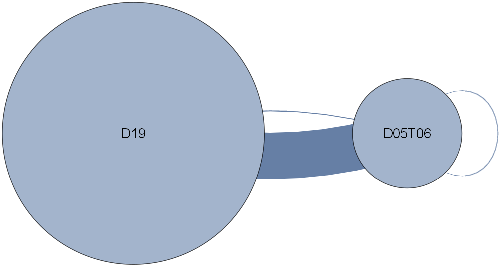}
	\caption{$\mathbf{S}=1.35676$.}
	\end{subfigure}
	\begin{subfigure}{0.475\textwidth}
	\includegraphics[width=\linewidth]{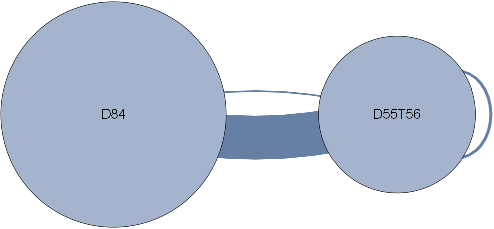}
	\caption{$\mathbf{S}=1.34825$.}
	\end{subfigure}
	\begin{subfigure}{0.475\textwidth}
	\includegraphics[width=\linewidth]{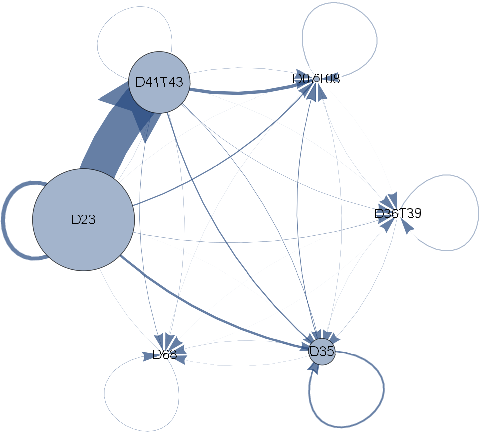}
	\caption{$\mathbf{S}=1.2768$.}
	\end{subfigure}
	\begin{subfigure}{0.475\textwidth}
	\includegraphics[width=\linewidth]{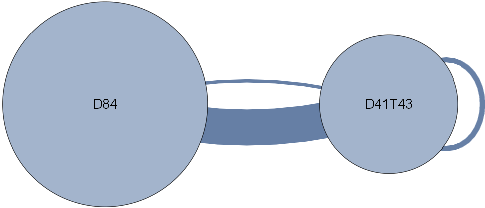}
	\caption{$\mathbf{S}=1.21107$.}
	\end{subfigure}
	\begin{subfigure}{0.475\textwidth}
	\includegraphics[width=\linewidth]{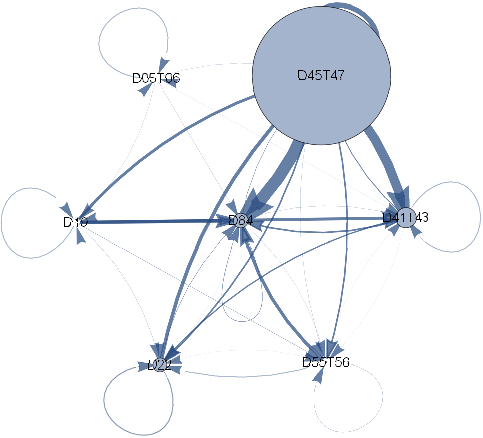}
	\caption{$\mathbf{S}=1.19601$.}
	\end{subfigure}
	\end{center}
	\caption{The fifteen overlapping agglomerative clusters of $2018$ Moroccan input-output table -- part~1.}
	\label{Fig7}
	\end{figure}

\begin{figure}[!htb]
	\begin{center}
	\begin{subfigure}{0.46\textwidth}
	\includegraphics[width=\linewidth]{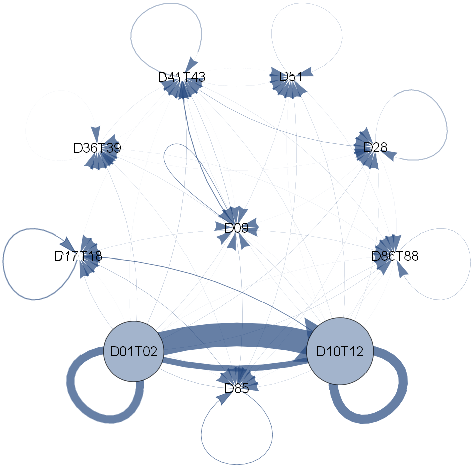}
	\caption{$\mathbf{S}=1.16461$.}
	\end{subfigure}
	\begin{subfigure}{0.46\textwidth}
	\includegraphics[width=\linewidth]{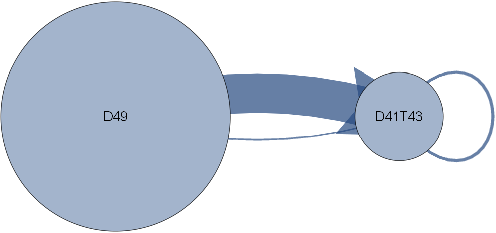}
	\caption{$\mathbf{S}=1.12018$.}
	\end{subfigure}
	\begin{subfigure}{0.46\textwidth}
	\includegraphics[width=\linewidth]{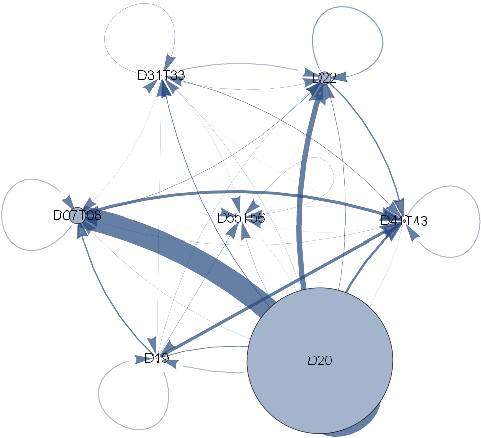}
	\caption{$\mathbf{S}=1.11361$.}
	\end{subfigure}
	\begin{subfigure}{0.46\textwidth}
	\includegraphics[width=\linewidth]{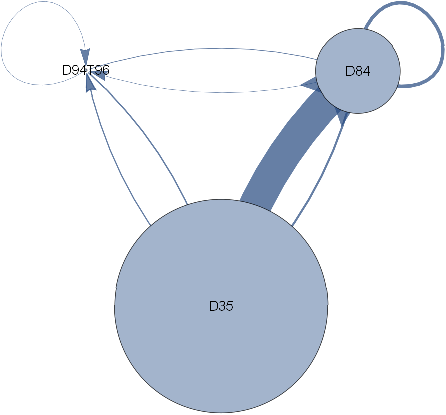}
	\caption{$\mathbf{S}=1.10433$.}
	\end{subfigure}
	\begin{subfigure}{0.46\textwidth}
	\includegraphics[width=\linewidth]{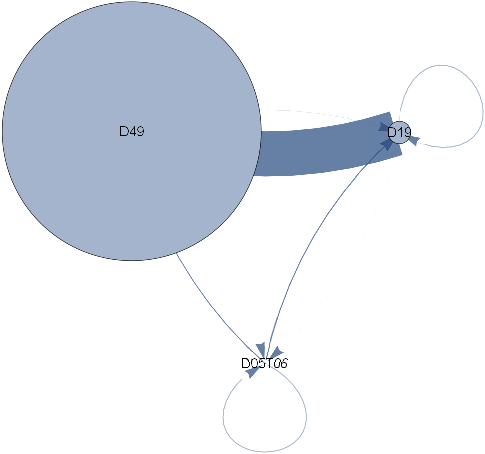}
	\caption{$\mathbf{S}=1.06336$.}
	\end{subfigure}
	\begin{subfigure}{0.46\textwidth}
	\includegraphics[width=\linewidth]{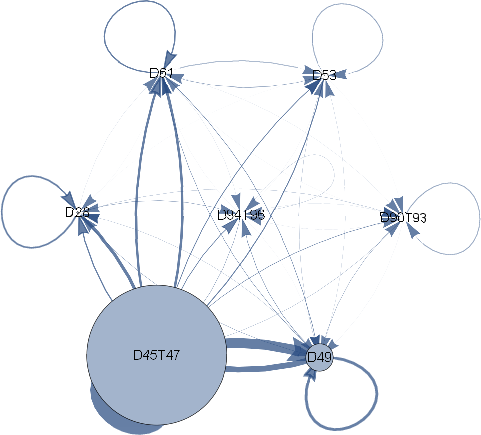}
	\caption{$\mathbf{S}=1.05842$.}
	\end{subfigure}
	\end{center}
	\caption{The fifteen overlapping agglomerative clusters of $2018$ Moroccan input-output table -- part~2.}
	\label{Fig8}
	\end{figure}

\clearpage

\begin{figure}[!htb]
	\begin{center}
	\begin{subfigure}{0.475\textwidth}
	\includegraphics[width=\linewidth]{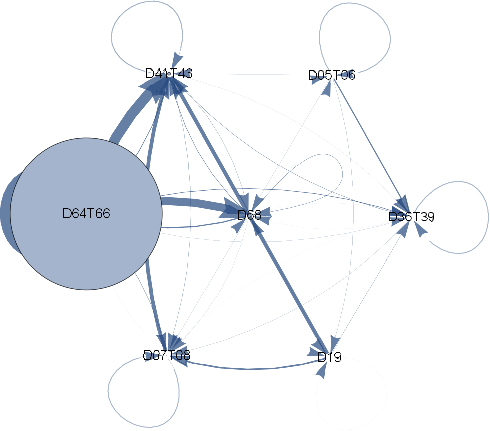}
	\caption{$\mathbf{S}=1.02491$.}
	\end{subfigure}
	\begin{subfigure}{0.475\textwidth}
	\includegraphics[width=\linewidth]{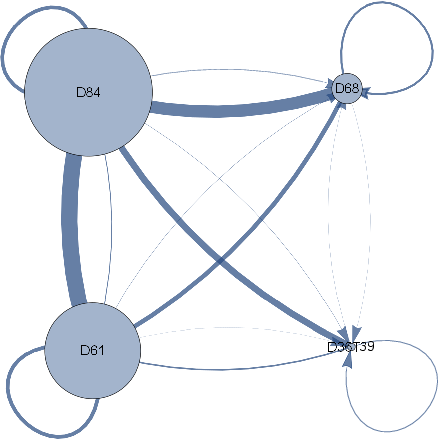}
	\caption{$\mathbf{S}=1.01399$.}
	\end{subfigure}
	\begin{subfigure}{0.475\textwidth}
	\includegraphics[width=\linewidth]{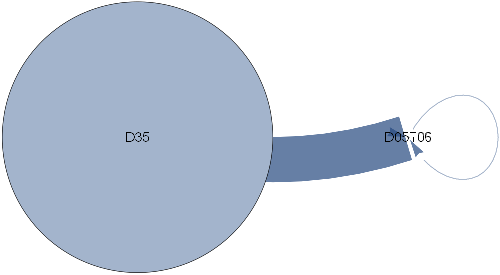}
	\caption{$\mathbf{S}=1.00407$.}
	\end{subfigure}
	\end{center}
	\caption{The fifteen overlapping agglomerative clusters of $2018$ Moroccan input-output table -- part~3.}
	\label{Fig9}
	\end{figure}

\vskip 1cm

A direct comparison between the non-overlapping and overlapping situations reveals the following observations:

\begin{itemize}
\item \textbf{Divisive algorithm:} The two algorithms yield nearly identical first four clusters (in the overlapping case, cluster number two includes a sixth pole D36T39: Water supply, sewerage, waste management and remediation activities, and cluster number four has a seventh pole D03: Fishing and aquaculture). The difference is observed in cluster number five, where the non-overlapping algorithm identifies a two-pole cluster related to mining and quarrying, while the overlapping algorithm identifies an eleven-pole cluster associated with public expenditures.

\item \textbf{Agglomerative algorithm:} The difference is much more significant, and additional clusters emerge when using overlapping clustering. The first three clusters are identical; however, the overlapping cluster number seven includes an additional pole (D41T43: Construction) compared to non-overlapping cluster number four. The non-overlapping cluster number five corresponds to the overlapping cluster number twelve, and the non-overlapping cluster number six is encompassed within overlapping cluster number thirteen.
\end{itemize}

\subsection{Discussion}

\hskip 0.5cm In $2018$, the structure of Moroccan interindustrial flows is characterized by high cyclicality. Compared to the countries in the panel of table  \ref{Tab3}, Morocco, along with Saudi Arabia, are the only countries with a spectral cyclicality greater than one, making them the countries with the highest interindustrial flows cyclicality. This reflects the fact that the Moroccan economy relies on a few propulsive clusters, leading to cyclical and interdependent industrial processes.

Three distinct propulsive clusters can be identified in the Moroccan economy. It is useful to draw an analogy with cyclical clusters and planetary systems in terms of influence, where a propulsive industry plays the role of the center of the system (akin to a planet or a star), while satellite industries orbit around it. These components can be described according to the following themes:

\begin{enumerate}
\item \textbf{Mining and quarrying} represented in figure \ref{Fig4}.(a), formed by a planet: D19-Coke and refined petroleum products; and a moon: D05T06-Mining and quarrying, energy producing products.

\item \textbf{Construction} represented in figure \ref{Fig4}.(b), characterized as planetary system of decreasing size (influence): D23-Other non-metallic mineral products; D41T43-Construction; D35-Electricity, gas, steam and air conditioning supply; D84-Public administration and defense, compulsory social security; D68-Real estate activities.

\item \textbf{Agri-food and related activities} represented in figure \ref{Fig4}.(c), formed by two planet-hosting stars: D10T12-Food products, beverages and tobacco; and D01T02-Agriculture, hunting, forestry; and seven planets: D17T18-Paper products and printing; D85-Education; D86T88-Human health and social work activities; D51-Air transport; D28-Machinery and equipment, nec; D36T39-Water supply, sewerage, waste management and remediation activities; D09-Mining support service activities.

\item \textbf{Logistics and transports} in figure \ref{Fig4}.(d), with a central star: D45T47-Wholesale and retail trade, repair of motor vehicles; two planets: D49-Land transport and transport via pipelines; D22-Rubber and plastics products; and three moons: D61-Telecommunications; D53-Postal and courier activities; D94T96-Other service activities.

\end{enumerate}

The study of intersecting clusters can reveal additional interesting productive components. Among them, one can identify the homogeneous ones:

\begin{enumerate}
\item \textbf{Energy products} represented in figure \ref{Fig7}.(f), represented by a star: D45T47: Wholesale and retail trade, repair of motor vehicles, three planets: D41T43: Construction, D84: Public administration and defense, compulsory social security, D22: Rubber and plastics products, and three satellites: D55T56: Accommodation and food service activities, D19: Coke and refined petroleum products, D05T06: Mining and quarrying, energy producing products.

\item \textbf{Manufacture of chemical products} represented in figure \ref{Fig8}.(c), represented by a star: D20: Chemical and chemical products, two planets: D07T08: Mining and quarrying, non-energy producing products, D22: Rubber and plastics products, and four satellites: D41T43: Construction, D31T33: Manufacturing nec, repair and installation of machinery and equipment, D55T56: Accommodation and food service activities, D19: Coke and refined petroleum products.

\end{enumerate}

The historical evolution of Moroccan interindustrial flows from $1995$ to $2018$, as depicted in figure \ref{Fig3}, reflects a structure that is clearly trending toward one characterized by strong cyclicality, marked by robust diffusion and interdependence. Morocco transitions from a single large cluster with weak cyclicality, encompassing nearly all sectors, to three distinct components with pronounced cyclicality. A detailed analysis is as follows:

\begin{enumerate}

\item The top propulsive industries, according to the spectral influence, consistently rank as poles: D10T12-Food products, beverages and tobacco; D01T02-Agriculture, hunting, forestry; D45T47-Wholesale and retail trade, repair of motor vehicles; D20-Chemical and chemical products. The weakest industries, in terms of spectral influence, in increasing order are: D97T98-Activities of households as employers, undifferentiated goods and services-producing activities of households for own use; D05T06-Mining and quarrying, energy producing products; D30-Other transport equipment.

\item According to the graphs depicting spectral influence evolution (see Appendix \ref{appendix4}), one can observe the structural evolution of the Moroccan economy, characterized by a process of concentration and complementarity in interindustrial flows. Notably, poles with increasing spectral influence can be observed: D01T02-Agriculture, hunting, forestry; D10T12-Food products, beverages and tobacco; D29-Motor vehicles, trailers and semi-trailers; D50-Water transport. This increasing influences absorb the influences of others poles like: D13T15-Textiles, textile products, leather and footwear; D16-Wood and products of wood and cork; D21-Pharmaceuticals, medicinal chemical and botanical products; D22-Rubber and plastics products; D23-Other non-metallic mineral products; D24-Basic metals; D25-Fabricated metal products; D26-Computer, electronic and optical equipment; D27-Electrical equipment; D28-Machinery and equipment, nec; D31T33-Manufacturing nec; repair and installation of machinery and equipment; D35-Electricity, gas, steam and air conditioning supply; D41T43-Construction; D45T47-Wholesale and retail trade; repair of motor vehicles; D53-Postal and courier activities; D55T56-Accommodation and food service activities; D61-Telecommunications; D68-Real estate activities.
\end{enumerate}

\section{Conclusion}

\hskip 0.5cm In summary, the novel concept of spectral influence acts as a local centrality measure for graph vertices, viewed through a diffusion lens. This measure provides both a local assessment and a holistic view of cyclic patterns across the entire graph. Leveraging this global measure, one can develop clustering algorithms aimed at identifying components with high cyclicality.

The application of this new technique to input-output analysis sheds new light to interindustrial organization by highlighting interindustrial transformation units involving many industries and detecting propulsive industries, which are the driving force of the economy. In contrast to the classical Leontieff/Ghosh analysis, which assumes a flow-relative orientation (demand/supply), spectral cyclicality analysis examines absolute flows and traces bilateral influences at time $t$.

Cyclicality in graphs reflects many interesting features that could be very helpful in understanding social networks, such as interdependence (dominance, resilience, etc.), diffusion (ideas, diseases, economic shocks, etc.), feedback effects, and amplification (productivity, transformation, etc.). This new technique has numerous potential applications.

\clearpage

\section{Appendix}

\subsection{Industrial poles of the OECD harmonised national input-output tables \label{appendix1}}

{\small
\setlength{\tabcolsep}{10pt} 
\renewcommand{\arraystretch}{1.5}

\begin{longtable}{p{5cm}p{5cm}p{5cm}}
\textbf{D01T02}: Agriculture, hunting, forestry & \textbf{D03}: Fishing and aquaculture & \textbf{D05T06}: Mining and quarrying, energy producing products \\
\textbf{D07T08}: Mining and quarrying, non-energy producing products & \textbf{D09}: Mining support service activities & \textbf{D10T12}: Food products, beverages and tobacco\\
\textbf{D13T15}: Textiles, textile products, leather and footwear & \textbf{D16}: Wood and products of wood and cork & \textbf{D17T18}: Paper products and printing \\
\textbf{D19}: Coke and refined petroleum products & \textbf{D20}: Chemical and chemical products & \textbf{D21}: Pharmaceuticals, medicinal chemical and botanical products\\
\textbf{D22}: Rubber and plastics products & \textbf{D23}: Other non-metallic mineral products & \textbf{D24}: Basic metals \\
\textbf{D25}: Fabricated metal products & \textbf{D26}: Computer, electronic and optical equipment & \textbf{D27}: Electrical equipment \\
\textbf{D28}: Machinery and equipment, nec & \textbf{D29}: Motor vehicles, trailers and semi-trailers & \textbf{D30}: Other transport equipment \\
\textbf{D31T33}: Manufacturing nec; repair and installation of machinery and equipment & \textbf{D35}: Electricity, gas, steam and air conditioning supply & \textbf{D36T39}: Water supply; sewerage, waste management and remediation activities \\
\textbf{D41T43}: Construction & \textbf{D45T47}: Wholesale and retail trade; repair of motor vehicles & \textbf{D49}: Land transport and transport via pipelines \\
\textbf{D50}: Water transport & \textbf{D51}: Air transport & \textbf{D52}: Warehousing and support activities for transportation \\
\textbf{D53}: Postal and courier activities & \textbf{D55T56}: Accommodation and food service activities & \textbf{D58T60}: Publishing, audiovisual and broadcasting activities \\
\textbf{D61}: Telecommunications & \textbf{D62T63}: IT and other information services & \textbf{D64T66}: Financial and insurance activities \\
\textbf{D68}: Real estate activities & \textbf{D69T75}: Professional, scientific and technical activities & \textbf{D77T82}: Administrative and support services \\
\textbf{D84}: Public administration and defense; compulsory social security & \textbf{D85}: Education & \textbf{D86T88}: Human health and social work activities \\
\textbf{D90T93}: Arts, entertainment and recreation & \textbf{D94T96}: Other service activities & \textbf{D97T98}: Activities of households as employers; undifferentiated goods- and services-producing activities of households for own use\\
\end{longtable}
}

\clearpage

\subsection{The Evolution of Cyclicality clusters in the Moroccan industrial structure \label{appendix2}}

\vskip -0.25cm

\begin{table}[!htb]
\centering
\begin{tabular}{|c|c|c|c|c|c|c|c|c|c|c|c|c|c|c|c|}
\hline
 &  \multicolumn{3}{|c|}{$\mathbf{1995}$} &  \multicolumn{3}{|c|}{$\mathbf{2000}$} &  \multicolumn{3}{|c|}{$\mathbf{2005}$} \tabularnewline
\hline
\textbf{Industry} & $\mathcal{S}$ & \textbf{DC} & \textbf{AC} & $\mathcal{S}$ & \textbf{DC} & \textbf{AC} & $\mathcal{S}$ & \textbf{DC} & \textbf{AC} \tabularnewline
\hline
\textbf{D01T02} & $0.16774$ & $1$ & $\iota$ & $0.301$ & $1$ & $\iota$ & $0.33535$ & $1$ & $\iota$ \tabularnewline
\hline
\textbf{D03} & $0.0001$ & $\iota$ & $\iota$ & $0.00008$ & $4$ & $6$ & $0.00016$ & $3$ & $5$ \tabularnewline
\hline
\textbf{D05T06} & $4*10^{-7}$ & $1$ & $6$ & $4*10^{-7}$ & $1$ & $4$ & $6*10^{-7}$ & $\iota$ & $5$ \tabularnewline
\hline
\textbf{D07T08} & $0.00025$ & $1$ & $\iota$ & $0.00002$ & $2$ & $2$ & $6*10^{-6}$ & $3$ & $4$ \tabularnewline
\hline
\textbf{D09} & $0.00005$ & $1$ & $6$ & $0.00001$ & $1$ & $4$ & $0.00001$ & $1$ & $\iota$ \tabularnewline
\hline
\textbf{D10T12} & $0.16811$ & $1$ & $\iota$ & $0.30329$ & $1$ & $\iota$ & $0.38024$ & $1$ & $\iota$ \tabularnewline
\hline
\textbf{D13T15} & $0.00542$ & $\iota$ & $\iota$ & $0.00179$ & $\iota$ & $\iota$ & $0.00095$ & $\iota$ & $\iota$ \tabularnewline
\hline
\textbf{D16} & $0.00016$ & $\iota$ & $\iota$ & $0.0001$ & $\iota$ & $\iota$ & $0.00007$ & $3$ & $\iota$ \tabularnewline
\hline
\textbf{D17T18} & $0.0002$ & $1$ & $\iota$ & $0.00015$ & $\iota$ & $\iota$ & $0.00011$ & $\iota$ & $\iota$ \tabularnewline
\hline
\textbf{D19} & $0.00138$ & $1$ & $1$ & $0.00123$ & $1$ & $1$ & $0.00136$ & $1$ & $1$ \tabularnewline
\hline
\textbf{D20} & $0.00581$ & $1$ & $\iota$ & $0.00315$ & $1$ & $\iota$ & $0.00266$ & $3$ & $\iota$ \tabularnewline
\hline
\textbf{D21} & $0.00062$ & $1$ & $\iota$ & $0.00043$ & $1$ & $\iota$ & $0.00037$ & $3$ & $\iota$ \tabularnewline
\hline
\textbf{D22} & $0.00072$ & $\iota$ & $\iota$ & $0.00033$ & $\iota$ & $6$ & $0.00037$ & $1$ & $5$  \tabularnewline
\hline
\textbf{D23} & $0.0004$ & $2$ & $\iota$ & $0.00014$ & $\iota$ & $\iota$ & $0.00011$ & $3$ & $\iota$ \tabularnewline 
\hline
\textbf{D24} & $0.00018$ & $1$ & $4$ & $0.00007$ & $3$ & $4$ & $0.00009$ & $3$ & $4$  \tabularnewline
\hline
\textbf{D25} & $0.00038$ & $\iota$ & $4$ & $0.00021$ & $3$ & $4$ & $0.00015$ & $3$ & $4$ \tabularnewline
\hline
\textbf{D26} & $0.00005$ & $1$ & $\iota$ & $0.00001$ & $\iota$ & $\iota$ & $0.00001$ & $1$ & $\iota$ \tabularnewline
\hline
\textbf{D27} & $0.0004$ & $\iota$ & $\iota$ & $0.00016$ & $\iota$ & $\iota$ & $0.00012$ & $3$ & $\iota$ \tabularnewline
\hline
\textbf{D28} & $0.00007$ & $1$ & $\iota$ & $0.00003$ & $1$ & $\iota$ & $0.00003$ & $1$ & $\iota$ \tabularnewline
\hline
\textbf{D29} & $0.00005$ & $\iota$ & $\iota$ & $0.00002$ & $\iota$ & $\iota$ & $0.00002$ & $3$ & $\iota$ \tabularnewline
\hline
\textbf{D30} & $2*10^{-6}$ & $1$ & $\iota$ & $2*10^{-6}$ & $\iota$ & $\iota$ & $1*10^{-6}$ & $3$ & $\iota$ \tabularnewline
\hline
\textbf{D31T33} & $0.00033$ & $1$ & $\iota$ & $0.00013$ & $4$ & $\iota$ & $0.00008$ & $3$ & $5$ \tabularnewline
\hline
\textbf{D35} & $0.00077$ & $1$ & $1$ & $0.00029$ & $1$ & $\iota$ & $0.0002$ & $1$ & $\iota$ \tabularnewline
\hline
\textbf{D36T39} & $0.00001$ & $1$ & $5$ & $0.00001$ & $1$ & $2$ & $0.00001$ & $1$ & $\iota$  \tabularnewline
\hline
\textbf{D41T43} & $0.00115$ & $2$ & $2$ & $0.00036$ & $2$ & $2$ & $0.00028$ & $2$ & $2$ \tabularnewline
\hline
\textbf{D45T47} & $0.01764$ & $1$ & $\iota$  & $0.01603$ & $1$ & $6$ & $0.01614$ & $1$ & $5$ \tabularnewline
\hline
\textbf{D49} & $0.00051$ & $1$ & $2$  & $0.0004$ & $1$ & $1$ & $0.0005$ & $1$ & $1$  \tabularnewline
\hline
\textbf{D50} & $0.00003$ & $1$ & $\iota$  & $0.00001$ & $1$ & $6$ & $0.00002$ & $1$ & $2$ \tabularnewline
\hline
\textbf{D51} & $0.00005$ & $1$ & $\iota$  & $0.00005$ & $1$ & $\iota$ & $0.00006$ & $1$ & $\iota$ \tabularnewline
\hline
\textbf{D52} & $0.0001$ & $1$ & $\iota$  & $0.00007$ & $1$ & $\iota$ & $0.00009$ & $1$ & $\iota$ \tabularnewline
\hline
\textbf{D53} & $0.00002$ & $1$ & $\iota$ & $7*10^{-6}$ & $4$ & $6$ & $8*10^{-6}$ & $1$ & $5$ \tabularnewline
\hline
\textbf{D55T56} & $0.00279$ & $1$ & $3$ & $0.00172$ & $1$ & $3$ & $0.0013$ & $1$ & $3$  \tabularnewline
\hline
\textbf{D58T60} & $0.0001$ & $1$ & $\iota$ & $0.00007$ & $1$ & $\iota$ & $0.00006$ & $1$ & $\iota$ \tabularnewline
\hline
\textbf{D61} & $0.00015$ & $1$ & $\iota$ & $0.00006$ & $1$ & $6$ & $0.00006$ & $1$ & $\iota$ \tabularnewline
\hline
\textbf{D62T63} & $0.00001$ & $1$ & $\iota$ & $7*10^{-6}$ & $4$ & $\iota$ & $5*10^{-6}$ & $3$ & $\iota$ \tabularnewline
\hline
\textbf{D64T66} & $0.00019$ & $1$ & $\iota$ & $0.00041$ & $1$ & $\iota$ & $0.00039$ & $1$ & $\iota$ \tabularnewline
\hline
\textbf{D68} & $0.00015$ & $1$ & $2$ & $0.00008$ & $4$ & $5$ & $0.00006$ & $2$ & $2$ \tabularnewline
\hline
\textbf{D69T75} & $0.00011$ & $1$ & $\iota$ & $0.0001$ & $\iota$ & $\iota$ & $0.0001$ & $3$ & $\iota$ \tabularnewline
\hline
\textbf{D77T82} & $0.00005$ & $1$ & $\iota$ & $0.00003$ & $1$ & $\iota$ & $0.00003$ & $1$ & $\iota$ \tabularnewline
\hline
\textbf{D84} & $0.00038$ & $1$ & $3$ & $0.00041$ & $1$ & $3$ & $0.00022$ & $1$ & $3$ \tabularnewline
\hline
\textbf{D85} & $0.00023$ & $1$ & $\iota$ & $0.00023$ & $1$ & $\iota$ & $0.00012$ & $1$ & $\iota$ \tabularnewline
\hline
\textbf{D86T88} & $0.00005$ & $1$ & $\iota$ & $0.00006$ & $1$ & $\iota$ & $0.00003$ & $1$ & $2$ \tabularnewline
\hline
\textbf{D90T93} & $2*10^{-6}$ & $1$ & $\iota$ & $1*10^{-6}$ & $1$ & $3$ & $8*10^{-7}$ & $1$ & $\iota$ \tabularnewline
\hline
\textbf{D94T96} & $0.00001$ & $1$ & $5$ & $0.00001$ & $1$ & $5$ & $7*10^{-6}$ & $1$ & $3$  \tabularnewline
\hline
\textbf{D97T98} & $0$ & $\iota$ & $\iota$ & $0$ & $\iota$ & $\iota$ & $0$ & $\iota$ & $\iota$ \tabularnewline
\hline
\end{tabular}
\captionof{table}{Evolution of interindustrial clusters - part 1.}
\label{Tab4}
\end{table}

\clearpage

\begin{table}[!htb]
\centering
\begin{tabular}{|c|c|c|c|c|c|c|c|c|c|}
\hline
 &  \multicolumn{3}{|c|}{$\mathbf{2010}$} &  \multicolumn{3}{|c|}{$\mathbf{2015}$} &  \multicolumn{3}{|c|}{$\mathbf{2018}$} \tabularnewline
\hline
\textbf{Industry} & $\mathcal{S}$ & \textbf{DC} & \textbf{AC} & $\mathcal{S}$ & \textbf{DC} & \textbf{AC} & $\mathcal{S}$ & \textbf{DC} & \textbf{AC} \tabularnewline
\hline
\textbf{D01T02} & $0.54341$ & $1$ & $4$ & $0.54843$ & $1$ & $4$ & $0.53322$ & $1$ & $4$ \tabularnewline
\hline
\textbf{D03} & $0.00008$ & $2$ & $\iota$ & $0.00011$ & $\iota$ & $\iota$ & $0.00007$ & $\iota$ & $\iota$ \tabularnewline
\hline
\textbf{D05T06} & $2*10^{-7}$ & $2$ & $5$ & $4*10^{-8}$ & $2$ & $5$  & $1*10^{-8}$ & $4$ & $2$ \tabularnewline
\hline
\textbf{D07T08} & $0.00019$ & $2$ & $1$ & $0.00008$ & $4$ & $1$ & $0.00008$& $5$ & $1$ \tabularnewline
\hline
\textbf{D09} & $7*10^{-6}$ & $1$ & $4$ & $4*10^{-6}$ & $1$ & $4$  & $3*10^{-6}$ & $1$ & $4$ \tabularnewline
\hline
\textbf{D10T12} & $0.52652$ & $1$ & $4$ & $0.58081$ & $1$ & $4$ & $0.58233$ & $1$ & $4$ \tabularnewline
\hline
\textbf{D13T15} & $0.00026$ & $\iota$ & $\iota$ & $0.00018$ & $\iota$ & $\iota$ & $0.00015$ & $\iota$ & $\iota$ \tabularnewline
\hline
\textbf{D16} & $0.00005$ & $\iota$ & $\iota$ & $0.00003$ & $\iota$ & $\iota$ & $0.00003$ & $\iota$ & $\iota$ \tabularnewline
\hline
\textbf{D17T18} & $0.00016$ & $1$ & $4$ & $0.00014$ & $1$ & $4$ & $0.00013$ & $1$ & $4$ \tabularnewline
\hline
\textbf{D19} & $0.00016$ & $1$ & $2$ & $0.00005$ & $1$ & $2$ & $7*10^{-7}$ & $4$ & $2$ \tabularnewline
\hline
\textbf{D20} & $0.00298$ & $\iota$ & $\iota$ & $0.00229$ & $4$ & $\iota$ & $0.00293$ & $5$ & $\iota$ \tabularnewline
\hline
\textbf{D21} & $0.00032$ & $\iota$ & $\iota$ & $0.00028$ & $\iota$ & $\iota$ & $0.00022$ & $\iota$ & $\iota$ \tabularnewline
\hline
\textbf{D22} & $0.00028$ & $2$ & $5$ & $0.00019$ & $2$ & $5$ & $0.00018$ & $3$ & $5$ \tabularnewline
\hline
\textbf{D23} & $0.00013$ & $\iota$ & $\iota$ & $0.00007$ & $\iota$ & $\iota$  & $0.00005$ & $2$ & $\iota$ \tabularnewline 
\hline
\textbf{D24} & $0.00004$ & $\iota$ & $\iota$ & $0.00001$ & $\iota$ & $\iota$ & $0.00001$ & $\iota$ & $\iota$ \tabularnewline
\hline
\textbf{D25} & $0.0001$ & $\iota$ & $\iota$ & $0.00005$ & $\iota$ & $\iota$ & $0.00006$ & $\iota$ & $\iota$ \tabularnewline
\hline
\textbf{D26} & $5*10^{-6}$ & $\iota$ & $4$ & $3*10^{-6}$ & $\iota$ & $\iota$ & $3*10^{-6}$ & $\iota$ & $\iota$ \tabularnewline
\hline
\textbf{D27} & $0.0001$ & $\iota$ & $\iota$ & $0.00004$ & $\iota$ & $\iota$  & $0.00004$ & $\iota$ & $\iota$ \tabularnewline
\hline
\textbf{D28} & $0.00002$ & $1$ & $4$ & $6*10^{-6}$ & $1$ & $4$ & $8*10^{-6}$ & $1$ & $4$ \tabularnewline
\hline
\textbf{D29} & $0.00003$ & $\iota$ & $\iota$ & $0.00007$ & $\iota$ & $\iota$ & $0.00009$ & $\iota$ & $\iota$ \tabularnewline
\hline
\textbf{D30} & $2*10^{-6}$ & $\iota$ & $\iota$ & $2*10^{-6}$ & $\iota$ & $\iota$ & $1*10^{-6}$ & $\iota$ & $\iota$ \tabularnewline
\hline
\textbf{D31T33} & $0.00004$ & $\iota$ & $\iota$ & $0.00002$ & $\iota$ & $\iota$ & $0.00002$ & $\iota$ & $\iota$ \tabularnewline
\hline
\textbf{D35} & $0.00014$ & $1$ & $4$ & $0.00012$ & $\iota$ & $\iota$ & $0.00015$ & $2$ & $\iota$ \tabularnewline
\hline
\textbf{D36T39} & $3*10^{-6}$ & $1$ & $4$ & $3*10^{-6}$ & $1$ & $4$ & $4*10^{-6}$ & $1$ & $4$ \tabularnewline
\hline
\textbf{D41T43} & $0.00036$ & $1$ & $1$ & $0.0002$ & $2$ & $1$ & $0.0002$ & $2$ & $1$ \tabularnewline
\hline
\textbf{D45T47} & $0.01347$ & $2$ & $5$ & $0.01056$ & $2$ & $5$ & $0.01057$ & $3$ & $5$ \tabularnewline
\hline
\textbf{D49} & $0.00035$ & $2$ & $2$ & $0.00029$ & $\iota$ & $2$ & $0.00033$ & $3$ & $5$ \tabularnewline
\hline
\textbf{D50} & $0.00003$ & $1$ & $4$ & $0.00003$ & $\iota$ & $\iota$ & $0.00005$ & $\iota$ & $\iota$ \tabularnewline
\hline
\textbf{D51} & $0.00003$ & $2$ & $5$ & $0.00004$ & $2$ & $5$ & $0.00004$ & $1$ & $4$ \tabularnewline
\hline
\textbf{D52} & $0.00008$ & $\iota$ & $\iota$ & $0.00007$ & $\iota$ & $\iota$ & $0.00007$ & $\iota$ & $\iota$ \tabularnewline
\hline
\textbf{D53} & $4*10^{-6}$ & $2$ & $5$ & $2*10^{-6}$ & $2$ & $5$ & $1*10^{-6}$ & $3$ & $5$ \tabularnewline
\hline
\textbf{D55T56} & $0.00059$ & $\iota$ & $3$ & $0.0005$ & $\iota$ & $3$ & $0.00067$ & $\iota$ & $3$ \tabularnewline
\hline
\textbf{D58T60} & $0.00003$ & $\iota$ & $\iota$ & $0.00003$ & $\iota$ & $\iota$ & $0.00003$ & $\iota$ & $\iota$ \tabularnewline
\hline
\textbf{D61} & $0.00004$ & $\iota$ & $5$ & $0.00002$ & $\iota$ & $5$ & $0.00002$ & $3$ & $5$ \tabularnewline
\hline
\textbf{D62T63} & $2*10^{-6}$ & $\iota$ & $6$ & $3*10^{-6}$ & $3$ & $\iota$ & $3*10^{-6}$ & $\iota$ & $\iota$ \tabularnewline
\hline
\textbf{D64T66} & $0.00034$ & $\iota$ & $6$ & $0.00031$ & $3$ & $\iota$ & $0.00035$ & $\iota$ & $6$ \tabularnewline
\hline
\textbf{D68} & $0.00002$ & $2$ & $6$ & $0.00002$ & $3$ & $5$ & $0.00002$ & $2$ & $6$ \tabularnewline
\hline
\textbf{D69T75} & $0.00007$ & $\iota$ & $\iota$ & $0.00006$ & $\iota$ & $\iota$ & $0.00007$ & $\iota$ & $\iota$ \tabularnewline
\hline
\textbf{D77T82} & $0.00001$ & $\iota$ & $6$ & $0.00002$ & $3$ & $\iota$ & $0.00002$ & $\iota$ & $\iota$ \tabularnewline
\hline
\textbf{D84} & $0.00029$ & $\iota$ & $3$ & $0.00028$ & $\iota$ & $3$ & $0.00031$ & $2$ & $3$ \tabularnewline
\hline
\textbf{D85} & $0.00013$ & $1$ & $4$ & $0.0001$ & $1$ & $4$ & $0.00011$ & $1$ & $4$ \tabularnewline
\hline
\textbf{D86T88} & $0.00005$ & $1$ & $4$ & $0.00003$ & $1$ & $4$ & $0.00005$ & $1$ & $4$ \tabularnewline
\hline
\textbf{D90T93} & $1*10^{-6}$ & $2$ & $5$ & $1*10^{-6}$ & $\iota$ & $\iota$ & $2*10^{-6}$ & $\iota$ & $\iota$ \tabularnewline
\hline
\textbf{D94T96} & $4*10^{-6}$ & $\iota$ & $5$ & $4*10^{-6}$ & $\iota$ & $\iota$ & $4*10^{-6}$ & $3$ & $5$ \tabularnewline
\hline
\textbf{D97T98} & $0$ & $\iota$ & $\iota$ & $0$ & $\iota$ & $\iota$  & $0$ & $\iota$ & $\iota$ \tabularnewline
\hline
\end{tabular}
\captionof{table}{Evolution of interindustrial clusters - part 2.}
\label{Tab5}
\end{table}

\clearpage

\subsection{Key cyclicality components in the Moroccan industrial structure \label{appendix3}}

\subsubsection{non-overlapping techniques}

\begin{minipage}[t]{0.5\textwidth}
\textbf{Divisive clustering}:\\
\footnotesize{
\begin{enumerate}

\item \textbf{Component I}:

\begin{enumerate}
\item D19: Coke and refined petroleum products,
\item D05T06: Mining and quarrying, energy producing products.
\end{enumerate}

\item \textbf{Component II}:

\begin{enumerate}
\item D23: Other non-metallic mineral products,
\item D41T43: Construction, 
\item D35: Electricity, gas, steam and air conditioning supply,
\item D84: Public administration and defense, compulsory social security,
\item D68: Real estate activities.
\end{enumerate}

\item \textbf{Component III}:

\begin{enumerate}
\item D10T12: Food products, beverages and tobacco,
\item D01T02: Agriculture, hunting, forestry,
\item D17T18: Paper products and printing,
\item D85: Education,
\item D86T88: Human health and social work activities,
\item D51: Air transport,
\item D28: Machinery and equipment, nec,
\item D36T39: Water supply, sewerage, waste management and remediation activities,
\item D09: Mining support service activities.
\end{enumerate}

\item \textbf{Component IV}:

\begin{enumerate}
\item D45T47: Wholesale and retail trade, repair of motor vehicles,
\item D49: Land transport and transport via pipelines,
\item D22: Rubber and plastics products,
\item D61: Telecommunications,
\item D53: Postal and courier activities,
\item D94T96: Other service activities.
\end{enumerate}

\item \textbf{Component V}:

\begin{enumerate}
\item D20: Chemical and chemical products,
\item D07T08: Mining and quarrying, non-energy producing products.
\end{enumerate}

\end{enumerate}
}
\end{minipage}
\begin{minipage}[t]{0.5\textwidth}
\textbf{Agglomerative clustering}:\\
\footnotesize{
\begin{enumerate}

\item \textbf{Component I}:

\begin{enumerate}
\item D41T43: Construction,
\item D07T08: Mining and quarrying, non-energy producing products.
\end{enumerate}

\item \textbf{Component II}:

\begin{enumerate}
\item D19: Coke and refined petroleum products,
\item D05T06: Mining and quarrying, energy producing products.
\end{enumerate}

\item \textbf{Component III}:

\begin{enumerate}
\item D84: Public administration and defense, compulsory social security,
\item D55T56: Accommodation and food service activities.
\end{enumerate}

\item \textbf{Component IV}:

\begin{enumerate}
\item D10T12: Food products, beverages and tobacco,
\item D01T02: Agriculture, hunting, forestry,
\item D17T18: Paper products and printing,
\item D85: Education,
\item D86T88: Human health and social work activities,
\item D51: Air transport,
\item D28: Machinery and equipment, nec,
\item D36T39: Water supply, sewerage, waste management and remediation activities,
\item D09: Mining support service activities.
\end{enumerate}

\item \textbf{Component V}:

\begin{enumerate}
\item D45T47: Wholesale and retail trade, repair of motor vehicles,
\item D49: Land transport and transport via pipelines,
\item D22: Rubber and plastics products,
\item D61: Telecommunications,
\item D53: Postal and courier activities,
\item D94T96: Other service activities.
\end{enumerate}

\item \textbf{Component VI}:

\begin{enumerate}
\item D64T66: Financial and insurance activities, 
\item D68: Real estate activities.
\end{enumerate}

\end{enumerate}
}
\end{minipage}

\clearpage

\subsubsection{Overlapping techniques}

\textbf{Overlapping divisive clustering}:\\

\begin{minipage}[t]{0.5\textwidth}
\begin{enumerate}
\item \textbf{Component I}:
\begin{enumerate}
\item D19: Coke and refined petroleum products,
\item D05T06: Mining and quarrying, energy producing products.
\end{enumerate}
\item \textbf{Component II}:
\begin{enumerate}
\item D23: Other non-metallic mineral products,
\item D41T43: Construction, 
\item D35: Electricity, gas, steam and air conditioning supply,
\item D84: Public administration and defense, compulsory social security,
\item D68: Real estate activities,
\item D36T39: Water supply, sewerage, waste management and remediation activities.
\end{enumerate}
\item \textbf{Component III}:
\begin{enumerate}
\item D10T12: Food products, beverages and tobacco,
\item D01T02: Agriculture, hunting, forestry,
\item D17T18: Paper products and printing,
\item D85: Education,
\item D86T88: Human health and social work activities,
\item D51: Air transport,
\item D28: Machinery and equipment, nec,
\item D36T39: Water supply, sewerage, waste management and remediation activities,
\item D09: Mining support service activities.
\end{enumerate}
\end{enumerate}
\end{minipage}
\begin{minipage}[t]{0.5\textwidth}
\begin{enumerate}
\setcounter{enumi}{3}
\item \textbf{Component IV}:
\begin{enumerate}
\item D45T47: Wholesale and retail trade, repair of motor vehicles,
\item D49: Land transport and transport via pipelines,
\item D22: Rubber and plastics products,
\item D61: Telecommunications,
\item D53: Postal and courier activities,
\item D94T96: Other service activities,
\item D03: Fishing and aquaculture.
\end{enumerate}
\item \textbf{Component V}:
\begin{enumerate}
\item D45T47: Wholesale and retail trade, repair of motor vehicles,
\item D84: Public administration and defense, compulsory social security,
\item D35: Electricity, gas, steam and air conditioning supply,
\item D51: Air transport,
\item D77T82: Administrative and support services,
\item D68: Real estate activities,
\item D31T33: Manufacturing nec, repair and installation of machinery and equipment,
\item D36T39: Water supply; sewerage, waste management and remediation \
activities,
\item D30: Other transport equipment,
\item D62T63: IT and other information services,
\item D19: Coke and refined petroleum products.
\end{enumerate}
\end{enumerate}
\end{minipage}

\clearpage

\textbf{Overlapping agglomerative clustering}:\\

{\tiny
\begin{minipage}[t]{0.5\textwidth}
\begin{enumerate}
\item \textbf{Component I}:
\begin{enumerate}
\item D41T43: Construction,
\item D07T08: Mining and quarrying, non-energy producing products.
\end{enumerate}
\item \textbf{Component II}:
\begin{enumerate}
\item D19: Coke and refined petroleum products,
\item D05T06: Mining and quarrying, energy producing products.
\end{enumerate}
\item \textbf{Component III}:
\begin{enumerate}
\item D84: Public administration and defense, compulsory social security,
\item D55T56: Accommodation and food service activities.
\end{enumerate}
\item \textbf{Component IV}:
\begin{enumerate}
\item D23: Other non-metallic mineral products,
\item D41T43: Construction,
\item D35: Electricity, gas, steam and air conditioning supply,
\item D07T08: Mining and quarrying, non-energy producing products,
\item D68: Real estate activities,
\item D36T39: Water supply; sewerage, waste management and remediation activities.
\end{enumerate}
\item \textbf{Component V}:
\begin{enumerate}
\item D84: Public administration and defense, compulsory social security,
\item D41T43: Construction.
\end{enumerate}
\item \textbf{Component VI}:
\begin{enumerate}
\item D45T47: Wholesale and retail trade; repair of motor vehicles,
\item D41T43: Construction,
\item D84: Public administration and defense, compulsory social security,
\item D22: Rubber and plastics products,
\item D55T56: Accommodation and food service activities,
\item D19: Coke and refined petroleum products,
\item D05T06: Mining and quarrying, energy producing products.
\end{enumerate}
\item \textbf{Component VII}:
\begin{enumerate}
\item D10T12: Food products, beverages and tobacco,
\item D01T02: Agriculture, hunting, forestry,
\item D17T18: Paper products and printing,
\item D85: Education,
\item D41T43: Construction,
\item D86T88: Human health and social work activities,
\item D51: Air transport,
\item D28: Machinery and equipment, nec,
\item D36T39: Water supply, sewerage, waste management and remediation activities,
\item D09: Mining support service activities.
\end{enumerate}
\item \textbf{Component VIII}:
\begin{enumerate}
\item D49: Land transport and transport via pipelines,
\item D41T43: Construction.
\end{enumerate}
\end{enumerate}
\end{minipage}
\begin{minipage}[t]{0.5\textwidth}
\begin{enumerate}
\setcounter{enumi}{8}
\item \textbf{Component IX}:
\begin{enumerate}
\item D20: Chemical and chemical products,
\item D07T08: Mining and quarrying, non-energy producing products.
\item D22: Rubber and plastics products,
\item D41T43: Construction,
\item D31T33: Manufacturing nec, repair and installation of machinery and equipment,
\item D55T56: Accommodation and food service activities,
\item D19: Coke and refined petroleum products.
\end{enumerate}
\item \textbf{Component X}:
\begin{enumerate}
\item D35: Electricity, gas, steam and air conditioning supply,
\item D84: Public administration and defense, compulsory social security,
\item D94T96: Other service activities.
\end{enumerate}
\item \textbf{Component XI}:
\begin{enumerate}
\item D49: Land transport and transport via pipelines,
\item D19: Coke and refined petroleum products,
\item D05T06: Mining and quarrying, energy producing products.
\end{enumerate}
\item \textbf{Component XII}:
\begin{enumerate}
\item D45T47: Wholesale and retail trade; repair of motor vehicles,
\item D49: Land transport and transport via pipelines,
\item D61: Telecommunications,
\item D28: Machinery and equipment, nec,
\item D53: Postal and courier activities,
\item D94T96: Other service activities,
\item D90T93: Arts, entertainment and recreation.
\end{enumerate}
\item \textbf{Component XIII}:
\begin{enumerate}
\item D64T66: Financial and insurance activities,
\item D68: Real estate activities,
\item D41T43: Construction,
\item D07T08: Mining and quarrying, non-energy producing products,
\item D36T39: Water supply; sewerage, waste management and remediation \
activities,
\item D19: Coke and refined petroleum products,
\item D05T06: Mining and quarrying, energy producing products.
\end{enumerate}
\item \textbf{Component XIV}:
\begin{enumerate}
\item D84: Public administration and defense, compulsory social security,
\item D61: Telecommunications,
\item D68: Real estate activities,
\item D36T39: Water supply; sewerage, waste management and remediation activities.
\end{enumerate}
\item \textbf{Component XV}:
\begin{enumerate}
\item D35: Electricity, gas, steam and air conditioning supply,
\item D05T06: Mining and quarrying, energy producing products.
\end{enumerate}
\end{enumerate}
\end{minipage}
}

\clearpage

\subsection{Spectral influence evolution by industrial poles \label{appendix4}}

\begin{figure}[!htb]
	\begin{center}
	\begin{subfigure}{0.225\textwidth}
	\includegraphics[width=\linewidth]{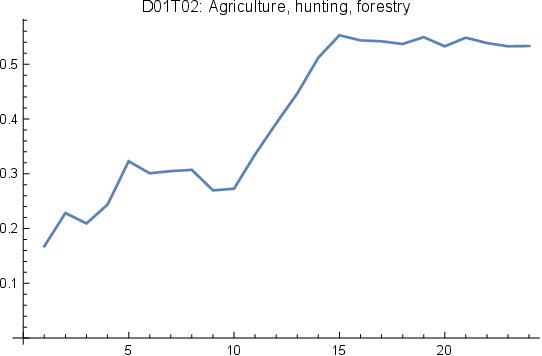}
	\caption*{D01T02.}
	\end{subfigure}
	\begin{subfigure}{0.225\textwidth}
	\includegraphics[width=\linewidth]{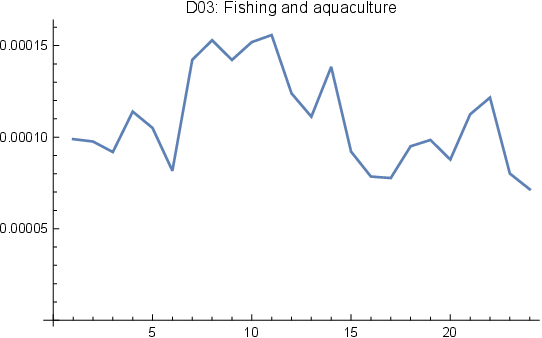}
	\caption*{D03.}
	\end{subfigure}
	\begin{subfigure}{0.2225\textwidth}
	\includegraphics[width=\linewidth]{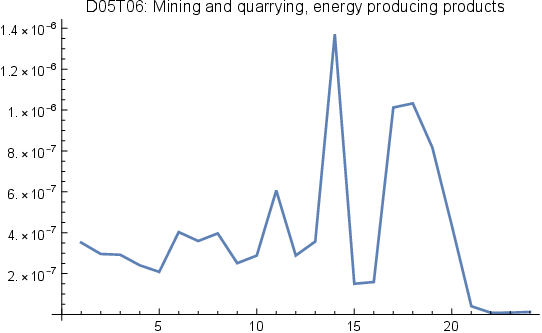}
	\caption*{D05T06.}
	\end{subfigure}
	\begin{subfigure}{0.225\textwidth}
	\includegraphics[width=\linewidth]{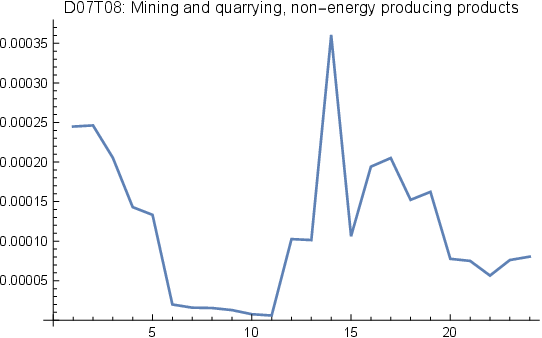}
	\caption*{D07T08.}
	\end{subfigure}
	\begin{subfigure}{0.225\textwidth}
	\includegraphics[width=\linewidth]{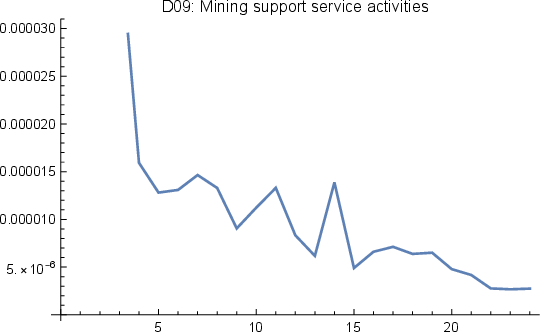}
	\caption*{D09.}
	\end{subfigure}
	\begin{subfigure}{0.225\textwidth}
	\includegraphics[width=\linewidth]{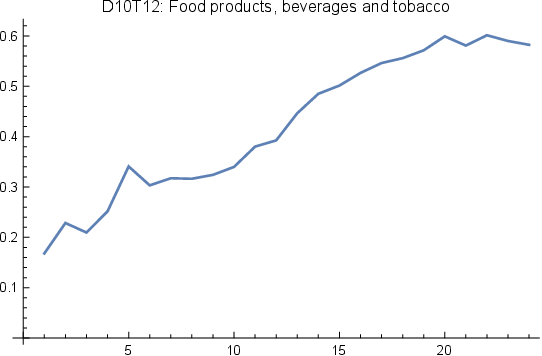}
	\caption*{D10T12.}
	\end{subfigure}
	\begin{subfigure}{0.2225\textwidth}
	\includegraphics[width=\linewidth]{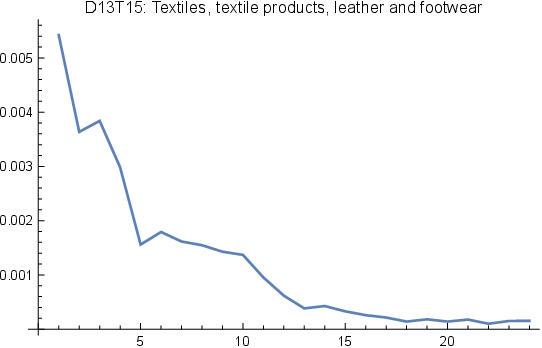}
	\caption*{D13T15.}
	\end{subfigure}
	\begin{subfigure}{0.225\textwidth}
	\includegraphics[width=\linewidth]{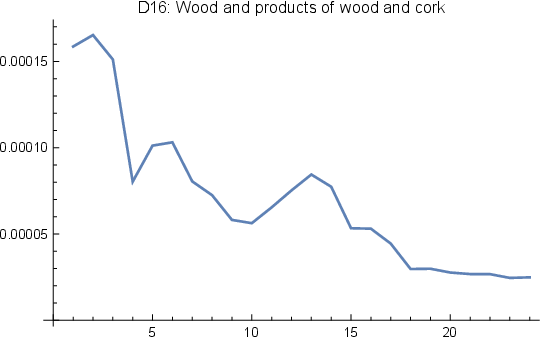}
	\caption*{D16.}
	\end{subfigure}
	\begin{subfigure}{0.225\textwidth}
	\includegraphics[width=\linewidth]{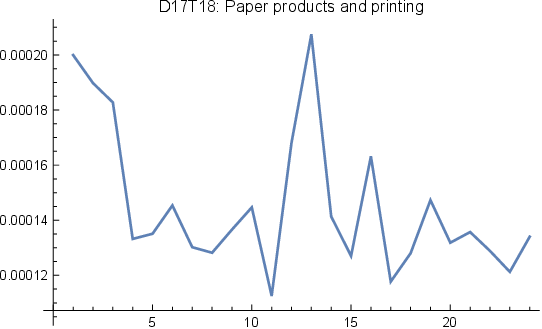}
	\caption*{D17T18.}
	\end{subfigure}
	\begin{subfigure}{0.225\textwidth}
	\includegraphics[width=\linewidth]{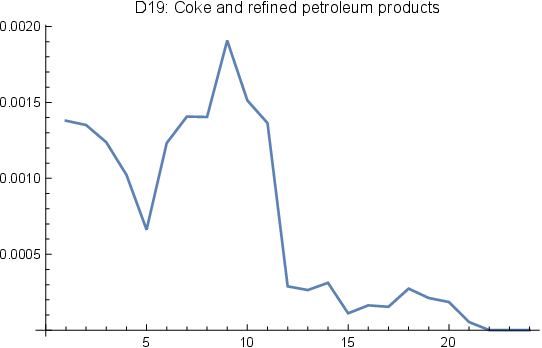}
	\caption*{D19.}
	\end{subfigure}
	\begin{subfigure}{0.2225\textwidth}
	\includegraphics[width=\linewidth]{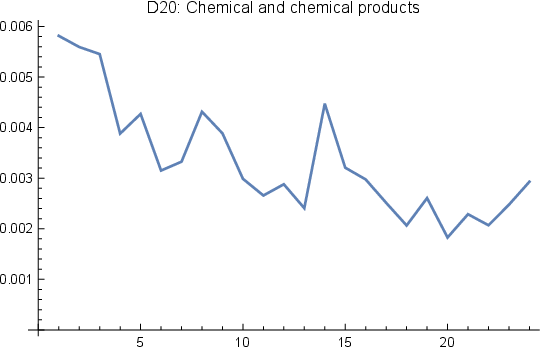}
	\caption*{D20.}
	\end{subfigure}
	\begin{subfigure}{0.225\textwidth}
	\includegraphics[width=\linewidth]{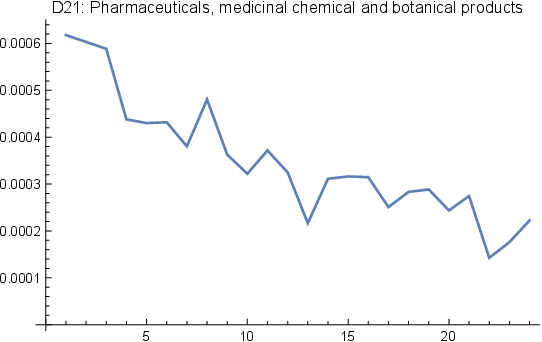}
	\caption*{D21.}
	\end{subfigure}
	\begin{subfigure}{0.225\textwidth}
	\includegraphics[width=\linewidth]{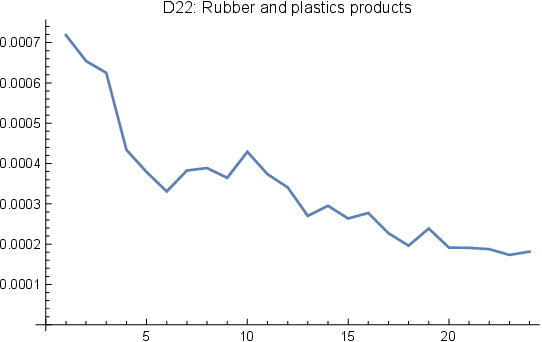}
	\caption*{D22.}
	\end{subfigure}
	\begin{subfigure}{0.225\textwidth}
	\includegraphics[width=\linewidth]{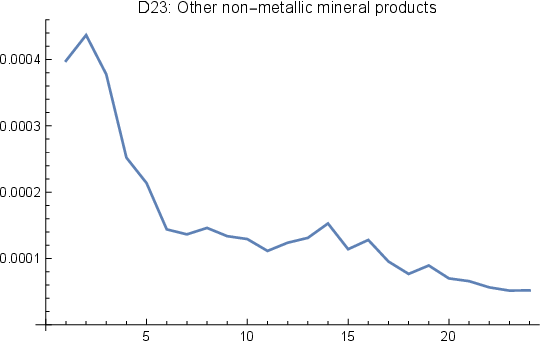}
	\caption*{D23.}
	\end{subfigure}
	\begin{subfigure}{0.2225\textwidth}
	\includegraphics[width=\linewidth]{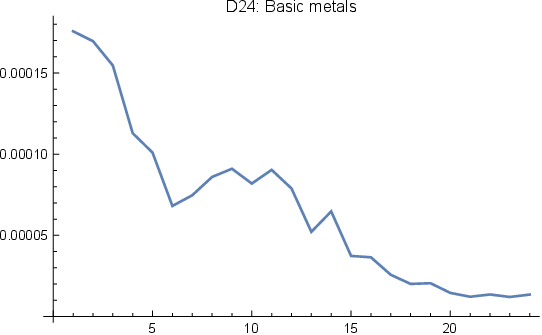}
	\caption*{D24.}
	\end{subfigure}
	\begin{subfigure}{0.225\textwidth}
	\includegraphics[width=\linewidth]{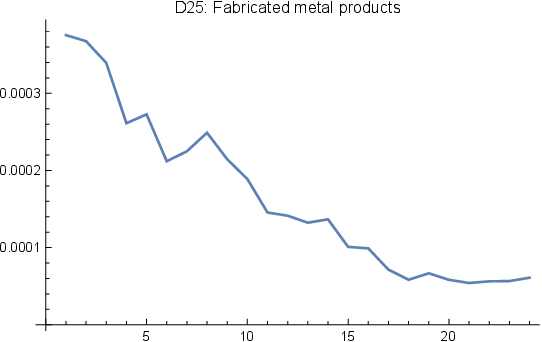}
	\caption*{D25.}
	\end{subfigure}
	\begin{subfigure}{0.225\textwidth}
	\includegraphics[width=\linewidth]{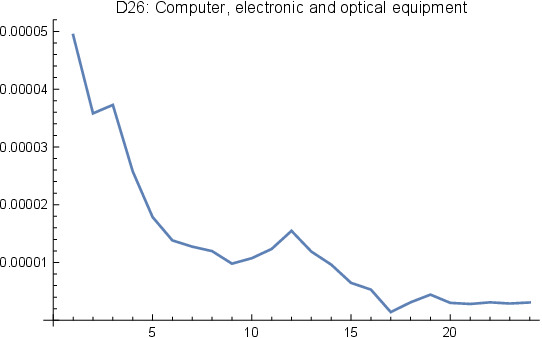}
	\caption*{D26.}
	\end{subfigure}
	\begin{subfigure}{0.225\textwidth}
	\includegraphics[width=\linewidth]{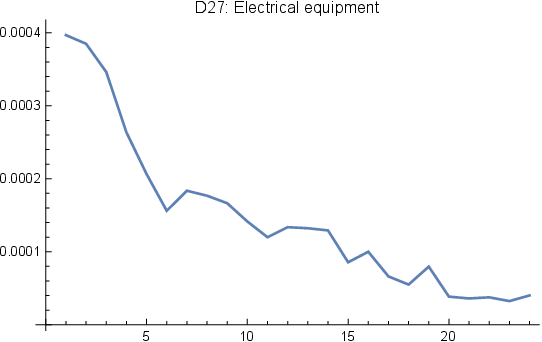}
	\caption*{D27.}
	\end{subfigure}
	\begin{subfigure}{0.2225\textwidth}
	\includegraphics[width=\linewidth]{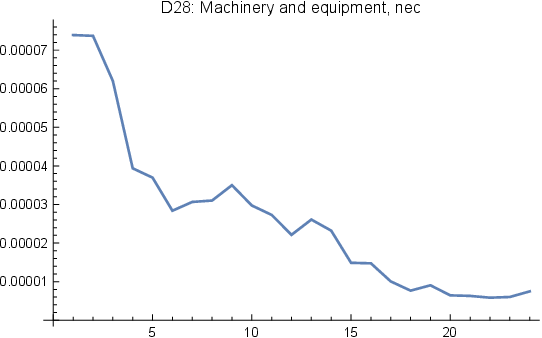}
	\caption*{D28.}
	\end{subfigure}
	\begin{subfigure}{0.225\textwidth}
	\includegraphics[width=\linewidth]{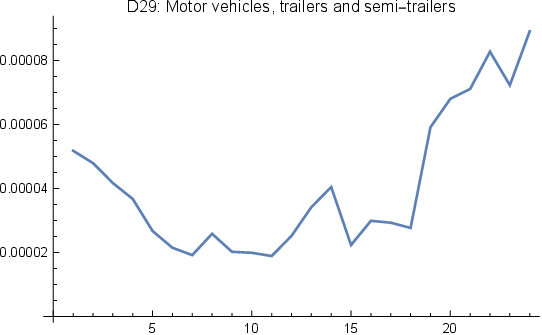}
	\caption*{D29.}
	\end{subfigure}
	\begin{subfigure}{0.225\textwidth}
	\includegraphics[width=\linewidth]{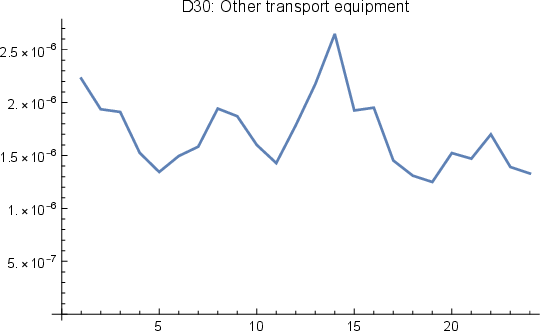}
	\caption*{D30.}
	\end{subfigure}
	\begin{subfigure}{0.225\textwidth}
	\includegraphics[width=\linewidth]{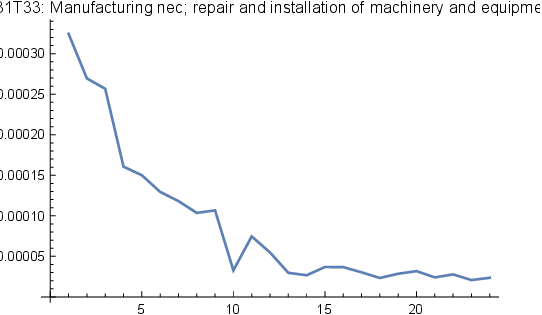}
	\caption*{D31T33.}
	\end{subfigure}
	\begin{subfigure}{0.2225\textwidth}
	\includegraphics[width=\linewidth]{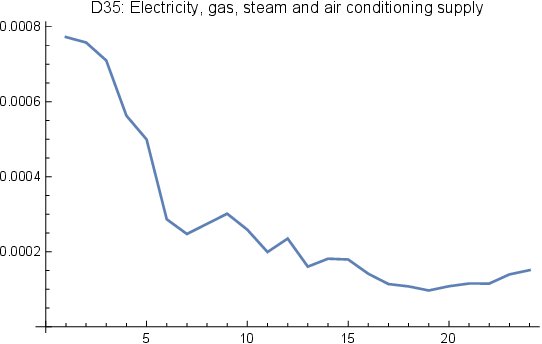}
	\caption*{D35.}
	\end{subfigure}
	\begin{subfigure}{0.225\textwidth}
	\includegraphics[width=\linewidth]{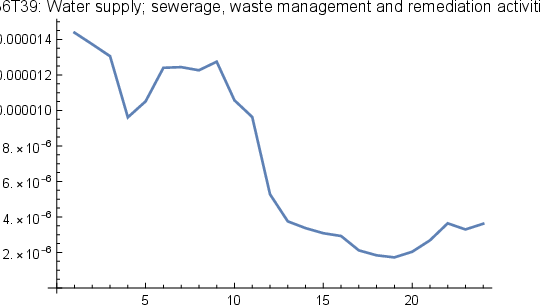}
	\caption*{D36T39.}
	\end{subfigure}
	\end{center}
	\caption{Evolution of the spectral radius of Moroccan industrial poles.\label{Fig10}}
	\end{figure}

\clearpage

\begin{figure}[!htb]
	\begin{center}
	\begin{subfigure}{0.225\textwidth}
	\includegraphics[width=\linewidth]{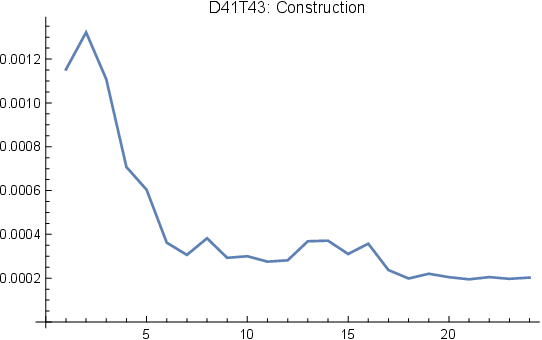}
	\caption*{D41T43.}
	\end{subfigure}
	\begin{subfigure}{0.225\textwidth}
	\includegraphics[width=\linewidth]{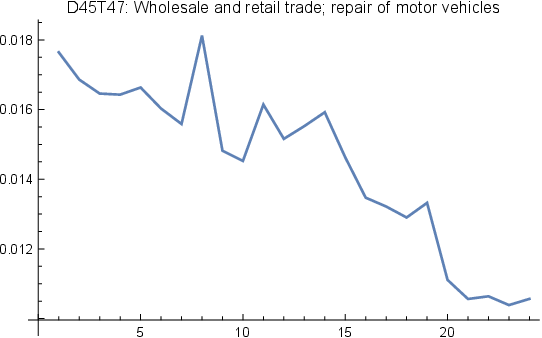}
	\caption*{D45T47.}
	\end{subfigure}
	\begin{subfigure}{0.2225\textwidth}
	\includegraphics[width=\linewidth]{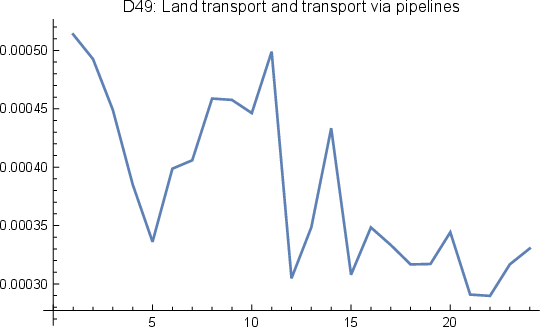}
	\caption*{D49.}
	\end{subfigure}
	\begin{subfigure}{0.225\textwidth}
	\includegraphics[width=\linewidth]{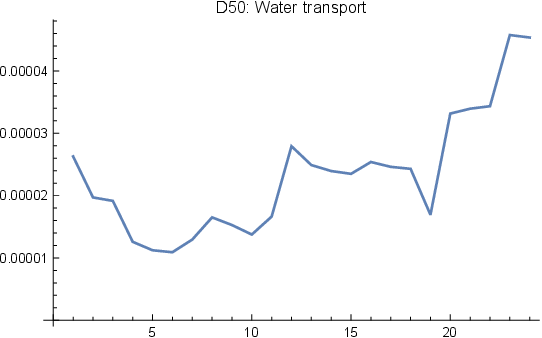}
	\caption*{D50.}
	\end{subfigure}
	\begin{subfigure}{0.225\textwidth}
	\includegraphics[width=\linewidth]{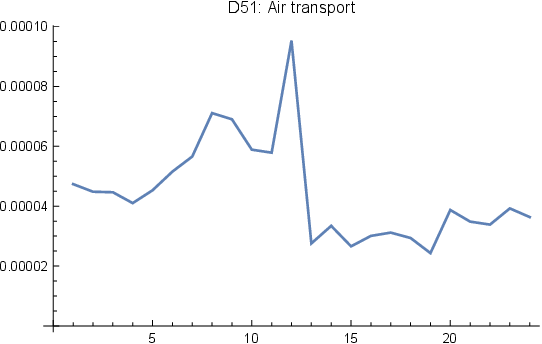}
	\caption*{D51.}
	\end{subfigure}
	\begin{subfigure}{0.225\textwidth}
	\includegraphics[width=\linewidth]{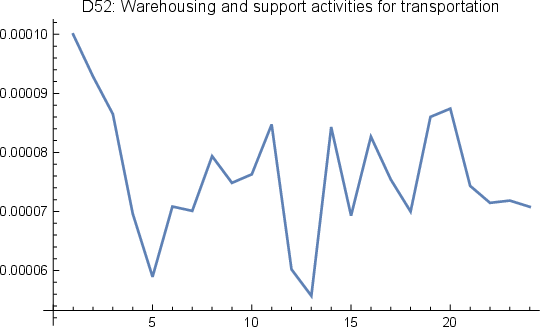}
	\caption*{D52.}
	\end{subfigure}
	\begin{subfigure}{0.2225\textwidth}
	\includegraphics[width=\linewidth]{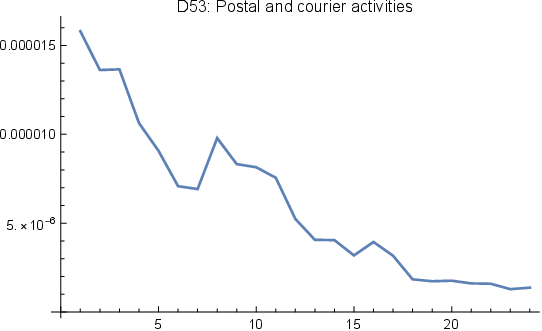}
	\caption*{D53.}
	\end{subfigure}
	\begin{subfigure}{0.225\textwidth}
	\includegraphics[width=\linewidth]{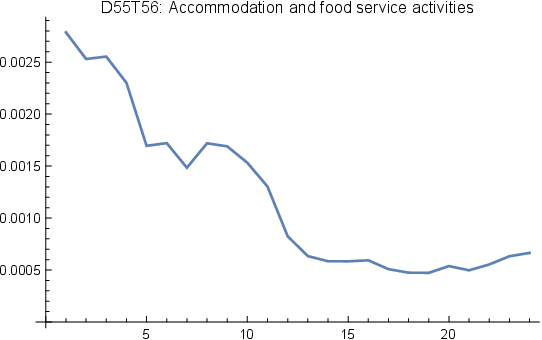}
	\caption*{D55T56.}
	\end{subfigure}
	\begin{subfigure}{0.225\textwidth}
	\includegraphics[width=\linewidth]{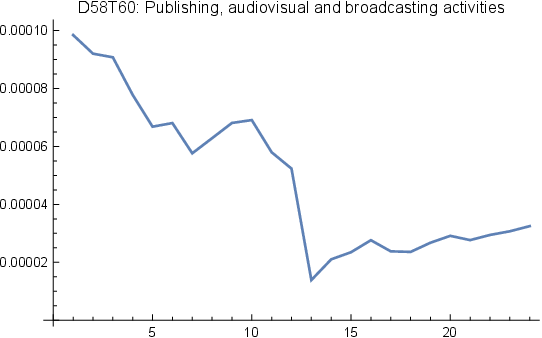}
	\caption*{D58T60.}
	\end{subfigure}
	\begin{subfigure}{0.225\textwidth}
	\includegraphics[width=\linewidth]{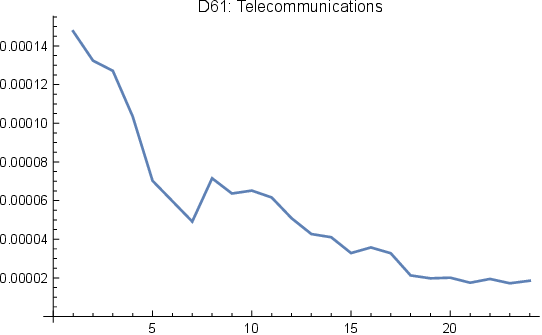}
	\caption*{D61.}
	\end{subfigure}
	\begin{subfigure}{0.2225\textwidth}
	\includegraphics[width=\linewidth]{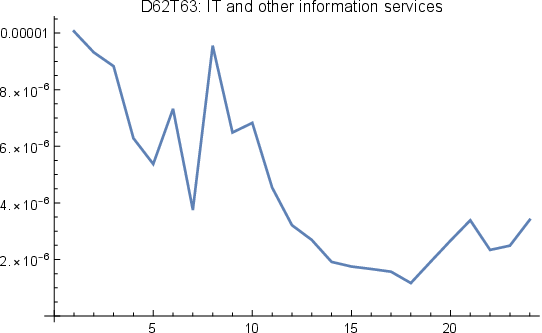}
	\caption*{D62T63.}
	\end{subfigure}
	\begin{subfigure}{0.225\textwidth}
	\includegraphics[width=\linewidth]{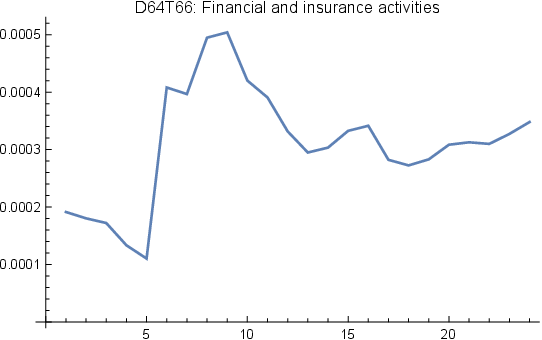}
	\caption*{D64T66.}
	\end{subfigure}
	\begin{subfigure}{0.225\textwidth}
	\includegraphics[width=\linewidth]{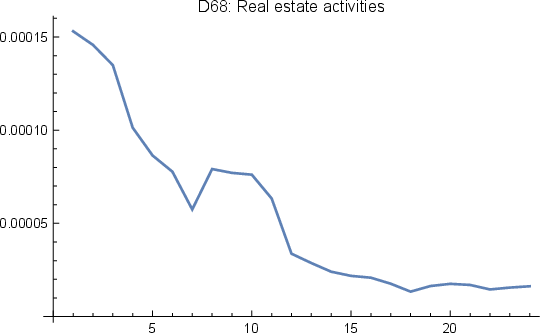}
	\caption*{D68.}
	\end{subfigure}
	\begin{subfigure}{0.225\textwidth}
	\includegraphics[width=\linewidth]{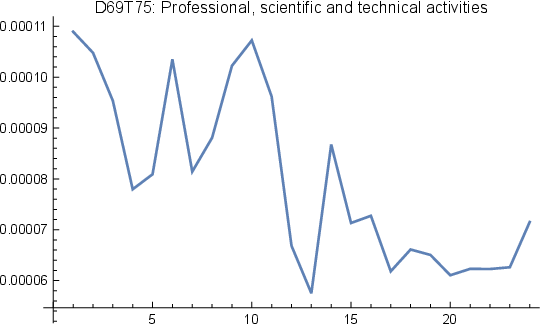}
	\caption*{D69T75.}
	\end{subfigure}
	\begin{subfigure}{0.2225\textwidth}
	\includegraphics[width=\linewidth]{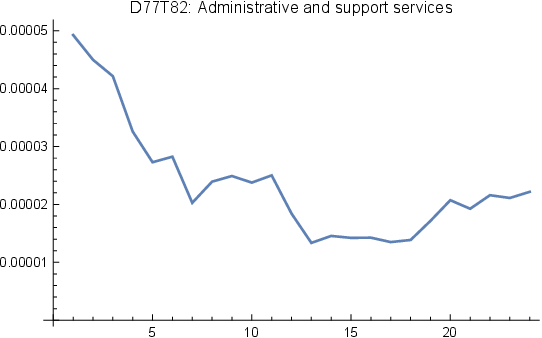}
	\caption*{D77T82.}
	\end{subfigure}
	\begin{subfigure}{0.225\textwidth}
	\includegraphics[width=\linewidth]{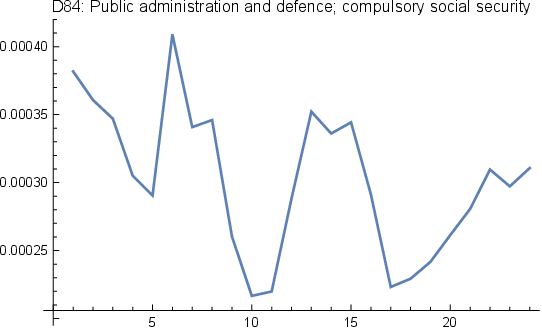}
	\caption*{D84.}
	\end{subfigure}
	\begin{subfigure}{0.225\textwidth}
	\includegraphics[width=\linewidth]{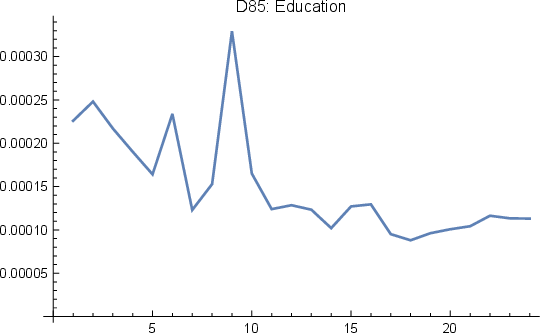}
	\caption*{D85.}
	\end{subfigure}
	\begin{subfigure}{0.225\textwidth}
	\includegraphics[width=\linewidth]{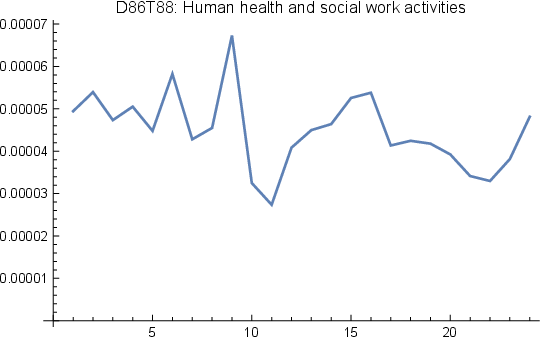}
	\caption*{D86T88.}
	\end{subfigure}
	\begin{subfigure}{0.225\textwidth}
	\includegraphics[width=\linewidth]{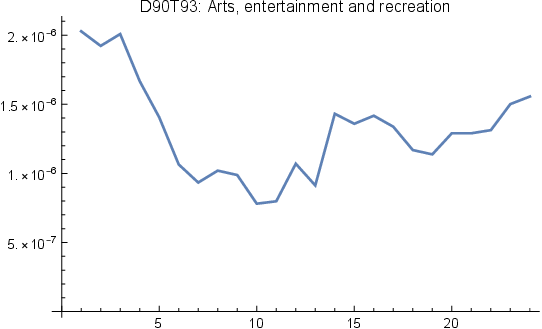}
	\caption*{D90T93.}
	\end{subfigure}
	\begin{subfigure}{0.225\textwidth}
	\includegraphics[width=\linewidth]{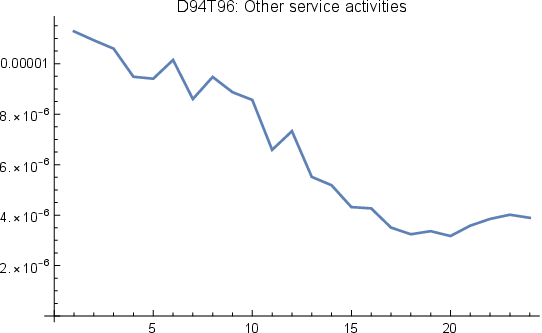}
	\caption*{D94T96.}
	\end{subfigure}
	\begin{subfigure}{0.225\textwidth}
	\includegraphics[width=\linewidth]{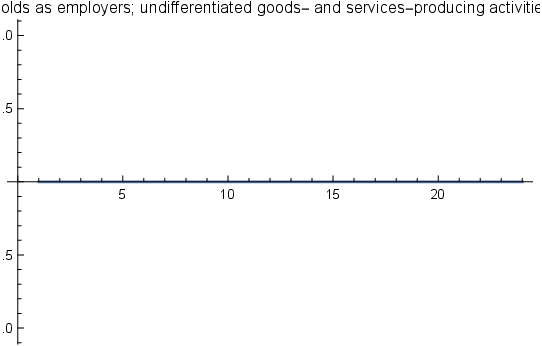}
	\caption*{D97T98.}
	\end{subfigure}
	\end{center}
	\caption{Evolution of the spectral radius of Moroccan industrial poles-bis.\label{Fig11}}
	\end{figure}

\bibliographystyle{alpha}
\bibliography{BibliographieGraph}

\end{document}